\documentclass[twocolumn]{aastex631}

\newcommand{\angstrom}{\mbox{\normalfont\AA}}
\usepackage[hyphens]{url}

\shorttitle{Dust Echo of AT\,2020mot}
\shortauthors{Newsome et al.}
\graphicspath{{./}}
\usepackage{amsmath}
\begin{document}

\title{Probing the Sub-Parsec Dust of a Supermassive Black Hole with the Tidal Disruption Event AT\,2020mot}

\newcommand{\LCO}{\affiliation{Las Cumbres Observatory, 6740 Cortona Drive, Suite 102, Goleta, CA 93117-5575, USA}}
\newcommand{\UCSB}{\affiliation{Department of Physics, University of California, Santa Barbara, CA 93106-9530, USA}}
\newcommand{\Iair}
{\affiliation{CIFAR Azrieli Global Scholars program, CIFAR, Toronto, Canada}}
\newcommand{\KITP}{\affiliation{Kavli Institute for Theoretical Physics, University of California, Santa Barbara, CA 93106-4030, USA}}
\newcommand{\UCD}{\affiliation{Department of Physics and Astronomy, University of California, Davis, 1 Shields Avenue, Davis, CA 95616-5270, USA}}
\newcommand{\WIS}{\affiliation{Department of Particle Physics and Astrophysics, Weizmann Institute of Science, 76100 Rehovot, Israel}}
\newcommand{\OKC}{\affiliation{Oskar Klein Centre, Department of Astronomy, Stockholm University, Albanova University Centre, SE-106 91 Stockholm, Sweden}}
\newcommand{\OAPD}{\affiliation{INAF -- Osservatorio Astronomico di Padova, Vicolo dell'Osservatorio 5, I-35122 Padova, Italy}}
\newcommand{\Caltech}{\affiliation{Cahill Center for Astronomy and Astrophysics, California Institute of Technology, Mail Code 249-17, Pasadena, CA 91125, USA}}
\newcommand{\GSFC}{\affiliation{Astrophysics Science Division, NASA Goddard Space Flight Center, Mail Code 661, Greenbelt, MD 20771, USA}}
\newcommand{\UMD}{\affiliation{Joint Space-Science Institute, University of Maryland, College Park, MD 20742, USA}}
\newcommand{\UCB}{\affiliation{Department of Astronomy, University of California, Berkeley, CA 94720-3411, USA}}
\newcommand{\TTU}{\affiliation{Department of Physics, Texas Tech University, Box 41051, Lubbock, TX 79409-1051, USA}}
\newcommand{\STScI}{\affiliation{Space Telescope Science Institute, 3700 San Martin Drive, Baltimore, MD 21218-2410, USA}}
\newcommand{\UT}{\affiliation{University of Texas at Austin, 1 University Station C1400, Austin, TX 78712-0259, USA}}
\newcommand{\IoA}{\affiliation{Institute of Astronomy, University of Cambridge, Madingley Road, Cambridge CB3 0HA, UK}}
\newcommand{\QUB}{\affiliation{Astrophysics Research Centre, School of Mathematics and Physics, Queen's University Belfast, Belfast BT7 1NN, UK}}
\newcommand{\IPAC}{\affiliation{Spitzer Science Center, California Institute of Technology, Pasadena, CA 91125, USA}}
\newcommand{\JPL}{\affiliation{Jet Propulsion Laboratory, California Institute of Technology, 4800 Oak Grove Dr, Pasadena, CA 91109, USA}}
\newcommand{\Southampton}{\affiliation{Department of Physics and Astronomy, University of Southampton, Southampton SO17 1BJ, UK}}
\newcommand{\LANL}{\affiliation{Space and Remote Sensing, MS B244, Los Alamos National Laboratory, Los Alamos, NM 87545, USA}}
\newcommand{\Tsinghua}{\affiliation{Physics Department and Tsinghua Center for Astrophysics, Tsinghua University, Beijing, 100084, People's Republic of China}}
\newcommand{\NAOC}{\affiliation{National Astronomical Observatory of China, Chinese Academy of Sciences, Beijing, 100012, People's Republic of China}}
\newcommand{\Itagaki}{\affiliation{Itagaki Astronomical Observatory, Yamagata 990-2492, Japan}}
\newcommand{\Einstein}{\altaffiliation{Einstein Fellow}}
\newcommand{\Hubble}{\altaffiliation{Hubble Fellow}}
\newcommand{\CfA}{\affiliation{Center for Astrophysics \textbar{} Harvard \& Smithsonian, 60 Garden Street, Cambridge, MA 02138-1516, USA}}
\newcommand{\UA}{\affiliation{Steward Observatory, University of Arizona, 933 North Cherry Avenue, Tucson, AZ 85721-0065, USA}}
\newcommand{\MPIA}{\affiliation{Max-Planck-Institut f\"ur Astrophysik, Karl-Schwarzschild-Stra\ss{}e 1, D-85748 Garching, Germany}}
\newcommand{\DSFP}{\altaffiliation{LSSTC Data Science Fellow}}
\newcommand{\HCO}{\affiliation{Harvard College Observatory, 60 Garden Street, Cambridge, MA 02138-1516, USA}}
\newcommand{\Carnegie}{\affiliation{Observatories of the Carnegie Institute for Science, 813 Santa Barbara Street, Pasadena, CA 91101-1232, USA}}
\newcommand{\TAU}{\affiliation{School of Physics and Astronomy, Tel Aviv University, Tel Aviv 69978, Israel}}
\newcommand{\Edinburgh}{\affiliation{Institute for Astronomy, University of Edinburgh, Royal Observatory, Blackford Hill EH9 3HJ, UK}}
\newcommand{\Birmingham}{\affiliation{Birmingham Institute for Gravitational Wave Astronomy and School of Physics and Astronomy, University of Birmingham, Birmingham B15 2TT, UK}}
\newcommand{\Bath}{\affiliation{Department of Physics, University of Bath, Claverton Down, Bath BA2 7AY, UK}}
\newcommand{\CTIO}{\affiliation{Cerro Tololo Inter-American Observatory, National Optical Astronomy Observatory, Casilla 603, La Serena, Chile}}
\newcommand{\Potsdam}{\affiliation{Institut f\"ur Physik und Astronomie, Universit\"at Potsdam, Haus 28, Karl-Liebknecht-Str. 24/25, D-14476 Potsdam-Golm, Germany}}
\newcommand{\INPE}{\affiliation{Instituto Nacional de Pesquisas Espaciais, Avenida dos Astronautas 1758, 12227-010, S\~ao Jos\'e dos Campos -- SP, Brazil}}
\newcommand{\UNC}{\affiliation{Department of Physics and Astronomy, University of North Carolina, 120 East Cameron Avenue, Chapel Hill, NC 27599, USA}}
\newcommand{\Ohio}{\affiliation{Astrophysical Institute, Department of Physics and Astronomy, 251B Clippinger Lab, Ohio University, Athens, OH 45701-2942, USA}}
\newcommand{\AAS}{\affiliation{American Astronomical Society, 1667 K~Street NW, Suite 800, Washington, DC 20006-1681, USA}}
\newcommand{\MMT}{\affiliation{MMT and Steward Observatories, University of Arizona, 933 North Cherry Avenue, Tucson, AZ 85721-0065, USA}}
\newcommand{\Geneva}{\affiliation{ISDC, Department of Astronomy, University of Geneva, Chemin d'\'Ecogia, 16 CH-1290 Versoix, Switzerland}}
\newcommand{\IUCAA}{\affiliation{Inter-University Center for Astronomy and Astrophysics, Post Bag 4, Ganeshkhind, Pune, Maharashtra 411007, India}}
\newcommand{\CMU}{\affiliation{Department of Physics, Carnegie Mellon University, 5000 Forbes Avenue, Pittsburgh, PA 15213-3815, USA}}
\newcommand{\NAOJ}{\affiliation{Division of Science, National Astronomical Observatory of Japan, 2-21-1 Osawa, Mitaka, Tokyo 181-8588, Japan}}
\newcommand{\IfA}{\affiliation{Institute for Astronomy, University of Hawai`i, 2680 Woodlawn Drive, Honolulu, HI 96822-1839, USA}}
\newcommand{\UCSC}{\affiliation{Department of Astronomy and Astrophysics, University of California, Santa Cruz, CA 95064-1077, USA}}
\newcommand{\Purdue}{\affiliation{Department of Physics and Astronomy, Purdue University, 525 Northwestern Avenue, West Lafayette, IN 47907-2036, USA}}
\newcommand{\Princeton}{\affiliation{Department of Astrophysical Sciences, Princeton University, 4 Ivy Lane, Princeton, NJ 08540-7219, USA}}
\newcommand{\Moore}{\affiliation{Gordon and Betty Moore Foundation, 1661 Page Mill Road, Palo Alto, CA 94304-1209, USA}}
\newcommand{\Durham}{\affiliation{Department of Physics, Durham University, South Road, Durham, DH1 3LE, UK}}
\newcommand{\JHU}{\affiliation{Department of Physics and Astronomy, The Johns Hopkins University, 3400 North Charles Street, Baltimore, MD 21218, USA}}
\newcommand{\Toronto}{\affiliation{David A.\ Dunlap Department of Astronomy and Astrophysics, University of Toronto,\\ 50 St.\ George Street, Toronto, Ontario, M5S 3H4 Canada}}
\newcommand{\Duke}{\affiliation{Department of Physics, Duke University, Campus Box 90305, Durham, NC 27708, USA}}
\newcommand{\NCU}{\affiliation{Graduate Institute of Astronomy, National Central University, 300 Jhongda Road, 32001 Jhongli, Taiwan}}
\newcommand{\Columbia}{\affiliation{Department of Physics and Columbia Astrophysics Laboratory, Columbia University, Pupin Hall, New York, NY 10027, USA}}
\newcommand{\Flatiron}{\affiliation{Center for Computational Astrophysics, Flatiron Institute, 162 5th Avenue, New York, NY 10010-5902, USA}}
\newcommand{\CIERA}{\affiliation{Center for Interdisciplinary Exploration and Research in Astrophysics and Department of Physics and Astronomy, \\Northwestern University, 1800 Sherman Avenue, 8th Floor, Evanston, IL 60201, USA}}
\newcommand{\GeminiObs}{\affiliation{Gemini Observatory, 670 North A`ohoku Place, Hilo, HI 96720-2700, USA}}
\newcommand{\Keck}{\affiliation{W.~M.~Keck Observatory, 65-1120 M\=amalahoa Highway, Kamuela, HI 96743-8431, USA}}
\newcommand{\UW}{\affiliation{Department of Astronomy, University of Washington, 3910 15th Avenue NE, Seattle, WA 98195-0002, USA}}
\newcommand{\DiRAC}{\altaffiliation{DiRAC Fellow}}
\newcommand{\USask}{\affiliation{Department of Physics \& Engineering Physics, University of Saskatchewan, 116 Science Place, Saskatoon, SK S7N 5E2, Canada}}
\newcommand{\Thacher}{\affiliation{Thacher School, 5025 Thacher Road, Ojai, CA 93023-8304, USA}}
\newcommand{\Rutgers}{\affiliation{Department of Physics and Astronomy, Rutgers, the State University of New Jersey,\\136 Frelinghuysen Road, Piscataway, NJ 08854-8019, USA}}
\newcommand{\FSU}{\affiliation{Department of Physics, Florida State University, 77 Chieftan Way, Tallahassee, FL 32306-4350, USA}}
\newcommand{\Melbourne}{\affiliation{School of Physics, The University of Melbourne, Parkville, VIC 3010, Australia}}
\newcommand{\ASTROthreeD}{\affiliation{ARC Centre of Excellence for All Sky Astrophysics in 3 Dimensions (ASTRO 3D)}}
\newcommand{\Stromlo}{\affiliation{Mt.\ Stromlo Observatory, The Research School of Astronomy and Astrophysics, Australian National University, ACT 2601, Australia}}
\newcommand{\NCPAS}{\affiliation{National Centre for the Public Awareness of Science, Australian National University, Canberra, ACT 2611, Australia}}
\newcommand{\TAMU}{\affiliation{Department of Physics and Astronomy, Texas A\&M University, 4242 TAMU, College Station, TX 77843, USA}}
\newcommand{\Mitchell}{\affiliation{George P.\ and Cynthia Woods Mitchell Institute for Fundamental Physics \& Astronomy, College Station, TX 77843, USA}}
\newcommand{\ESO}{\affiliation{European Southern Observatory, Alonso de C\'ordova 3107, Casilla 19, Santiago, Chile}}
\newcommand{\ICE}{\affiliation{Institute of Space Sciences (ICE, CSIC), Campus UAB, Carrer
de Can Magrans, s/n, E-08193 Barcelona, Spain}}
\newcommand{\IEEC}{\affiliation{Institut d'Estudis Espacials de Catalunya, Gran Capit\`a, 2-4, Edifici Nexus, Desp.\ 201, E-08034 Barcelona, Spain}}
\newcommand{\Warwick}{\affiliation{Department of Physics, University of Warwick, Gibbet Hill Road, Coventry CV4 7AL, UK}}
\newcommand{\Macquarie}{\affiliation{School of Mathematical and Physical Sciences, Macquarie University, NSW 2109, Australia}}
\newcommand{\AAARC}{\affiliation{Astronomy, Astrophysics and Astrophotonics Research Centre, Macquarie University, Sydney, NSW 2109, Australia}}
\newcommand{\Capodimonte}{\affiliation{INAF -- Capodimonte Astronomical Observatory, Salita Moiariello 16, I-80131 Napoli, Italy}}
\newcommand{\INFNNapoli}{\affiliation{INFN -- Napoli, Strada Comunale Cinthia, I-80126 Napoli, Italy}}
\newcommand{\ICRANet}{\affiliation{ICRANet, Piazza della Repubblica 10, I-65122 Pescara, Italy}}

\author[0000-0001-9570-0584]{Megan Newsome}
\LCO\UCSB
\author[0000-0001-7090-4898]{Iair Arcavi}
\TAU\Iair
\author[0000-0003-4253-656X]{D. Andrew Howell}
\LCO\UCSB
\author[0000-0003-0035-6659]{Jamison Burke}
\LCO\UCSB
\author[0000-0002-7579-1105]{Yael Dgany}
\TAU
\author[0009-0007-8485-1281]{Sara Faris}
\TAU
\author[0000-0003-4914-5625]{Joseph Farah}
\LCO\UCSB
\author[0000-0002-1125-9187]{Daichi Hiramatsu}
\affiliation{Center for Astrophysics \textbar{} Harvard \& Smithsonian, 60 Garden Street, Cambridge, MA 02138-1516, USA}
\affiliation{The NSF AI Institute for Artificial Intelligence and Fundamental Interactions, USA}
\author[0000-0001-5807-7893]{Curtis McCully}
\LCO
\author[0000-0003-0209-9246]{Estefania Padilla-Gonzalez}
\LCO\UCSB
\author[0000-0002-7472-1279]{Craig Pellegrino}
\LCO\UCSB
\author[0000-0003-0794-5982]{Giacomo Terreran}
\LCO



\begin{abstract}

 AT 2020mot is a typical UV/optical tidal disruption event (TDE) with no radio or X-ray signatures in a quiescent host. We find an $i$-band excess and re-brightening along the decline of the light curve which could be due to two consecutive dust echoes from a TDE. We model our observations following \cite{VanVelzen2016} and find that the near-infrared light curve can be explained by concentric rings of thin dust within $\sim$0.1 parsecs of a $\sim$6 $\times 10^{6} M_{\odot}$ supermassive black hole (SMBH), among the smallest scales at which dust has been inferred near SMBHs. We find dust covering factors of order $f_{c}\leq 2\%$, much lower than found for dusty tori of active galactic nuclei. These results highlight the potential of TDEs for uncovering the environments around black holes when including near-infrared observations in high-cadence transient studies.

\end{abstract}



\section{Introduction} \label{sec:intro}

Tidal disruption events (TDEs) occur when a star's self-gravity is disturbed by the tidal forces of a supermassive black hole (SMBH), after which the infall of the stripped, stellar debris causes a bright flare that is often observable in optical bands by transient surveys. The evolution of TDE light curves will vary based on the SMBH's mass, the unlucky star's mass before its tidal unraveling, and the depth of the impact. Furthermore, the source of the UV/optical emission is debated, whether as a result of the debris producing stream shocks on its fallback journey toward the SMBH \citep{Piran2015}, or if the material has already completed that journey and the flare is a result of accretion emission being reprocessed into the optical regime \citep[][]{Roth2016, Jiang2016, Guillochon2014}. Therefore TDE observations are often heralded as windows into accretion mechanisms, jet formation, and feedback effects on local and large-scale galactic environments. However, these implications are not yet well explored. In just a decade, the number of observed TDEs has escalated quickly \citep[e.g.][]{Graham2019, VanVelzen21, Hammerstein2023, Yao2023}, with over 50 reported to the Transient Name Server\footnote{https://www.wis-tns.org/} in the last five years. The increase in discoveries allows for large-sample statistics on TDE rates and host galaxy dependencies. We can now single out exceptional observations that probe the physics of accretion and feedback on sub-parsec scales around quiescent galactic nuclei.

If the tidal disruption flare interacts with circumnuclear dust, the effect can be observed as time-dependent reddening that allows deduction of the dust's geometry and location, and such visible signatures may be concurrent with the initial flare. Interaction with dust, for example, can cause a ``dust echo" in which the dust is heated from the incident light and subsequently re-emits in the infrared. Reprocessing light curves have long been used as probes of dust in other energetic phenomena like active galactic nuclei \citep[AGN; e.g.][]{1987ApJ...320..537B} and supernovae \citep[e.g.][]{1983ApJ...274..175D, 2005MNRAS.357.1161P}. The time delay at which a reprocessed infrared light curve becomes visible depends on how close the dust is to the central flare, while the shape of the reprocessed light curve varies with the dust's geometry. TDEs offer the unique chance to reveal dust at sub-parsec scales to quiescent SMBHs.

Just over a dozen dust echoes have been reported for candidate TDEs \citep[e.g.][]{JiangN2016, Dou2016, VanVelzen2016, Jiang2017, Li2020, Stein2021, Jiang2021, Wang2022, Onori2022}, using mid-infrared (MIR) data from the AllWISE and NEOWISE releases of the \textit{Wide-field Infrared Survey Explorer} \citep{2010AJ....140.1868W, 2014ApJ...792...30M}, most finding 2-5 epochs of MIR brightening after optical discovery of each TDE between 100--200 days after optical peak \citep{VanVelzen21}. Only eight of these events are in quiescent galaxies, seven of which have accompanying X-ray data alongside UV/optical detections \citep{VanVelzen21, Jiang2021}. Two of these, ASASSN-14li and AT\,2019dsg, were detected in the radio as well \citep{Jiang2016, Stein2021}. Using the delay between the UV/optical peak and rebrightening in MIR, the distance between the flare and the nearby dust was found to be $\sim 0.1 - 0.8$ pc for the eight events, and these echoes in quiescent galaxies subsequently show dust covering factors (fraction of the accretion disk or SMBH environment that is obscured by dust) of $f_{c} \sim 0.01$, notably lower than $f_{c} \sim 0.2 - 0.5$  as found in active galactic nuclei \citep[AGN, ][]{VanVelzen21}.

AT 2020mot was discovered by the Zwicky Transient Facility (ZTF) on MJD 59014.39 (June 14 2020) as a nuclear transient \citep{2020TNSTR1791....1F}, later classified by the Global Supernova Project as a TDE \citep{2020TNSCR2478....1H} given its central location to the host galaxy at (RA, Dec) = (00:31:13.57, +85:00:31.9) and broad He II emission feature. \citet{Hammerstein2023} include AT 2020mot in their sample of 30 TDEs, classifying it as a ``TDE-H+He" from observed broad H$\alpha$ and He II. \citet{Liodakis2022} also report polarized emission from AT 2020mot that is the highest measured for a TDE without a jet, with radio upper-limits of $27 \mu$Jy at 15 GHz.

Here we present data from Las Cumbres Observatory which reveals that AT 2020mot was still emitting significant light in the $i$-band as of +800 days from optical peak. We explain this infrared excess in an otherwise normal TDE as the result of a dust echo, in which dust near the SMBH is physically thin and close enough to respond promptly (within months) to the flare, heating up and re-emitting light in the near-infrared.

\section{Observations and Data Processing} \label{sec:style}

Throughout this work, we use observations taken by Las Cumbres Observatory, the Zwicky Transient Facility \citep[ZTF;][]{2019PASP..131a8002B}, the Niel Gehrels Swift Observatory \citep[Swift;][]{2004ApJ...611.1005G}, and the Wide-field Infrared Survey Explorer \citep[WISE;][]{2010AJ....140.1868W}. All phases are expressed in observer-frame throughout this work.

\subsection{Optical Photometry and Spectroscopy}

\begin{figure}[t]
\centering
    \includegraphics[scale=0.64]{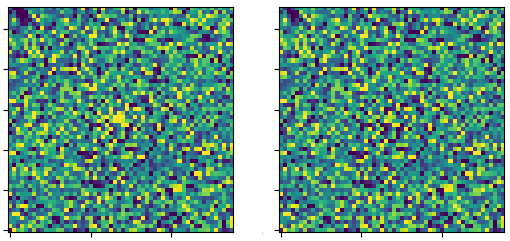}
    \caption{AT 2020mot in the $i$-band taken by Las Cumbres on +788 days from optical peak (MJD = 59078.38) and subtracted with archival PS1 images. Left: The PS1-subtracted image centered on the TDE. Right: The residual after subtracting off the PSF. The subtraction and residual show we still detect emission from the transient in the $i$-band +788 days after optical peak.}
    \label{fig:ps1_residual}
\end{figure}

We obtained $BgVri$-band images with the Sinistro cameras on Las Cumbres' 1.0m telescopes, performed image subtraction, and extracted point-spread-function (PSF) photometry with the \texttt{lcogtsnpipe} pipeline\footnote{\url{https://github.com/LCOGT/lcogtsnpipe}} \citep{Valenti16}. $BV$-band photometry was calibrated to Vega magnitudes and $gri$-band photometry was calibrated to the AB magnitude system. Zeropoints for all bands were calculated from the magnitudes of field stars as listed in the AAVSO Photometry All-Sky Survey \citep{2009AAS...21440702H}.

\begin{figure*}[t!]
    \centering
    \includegraphics[scale=0.75]{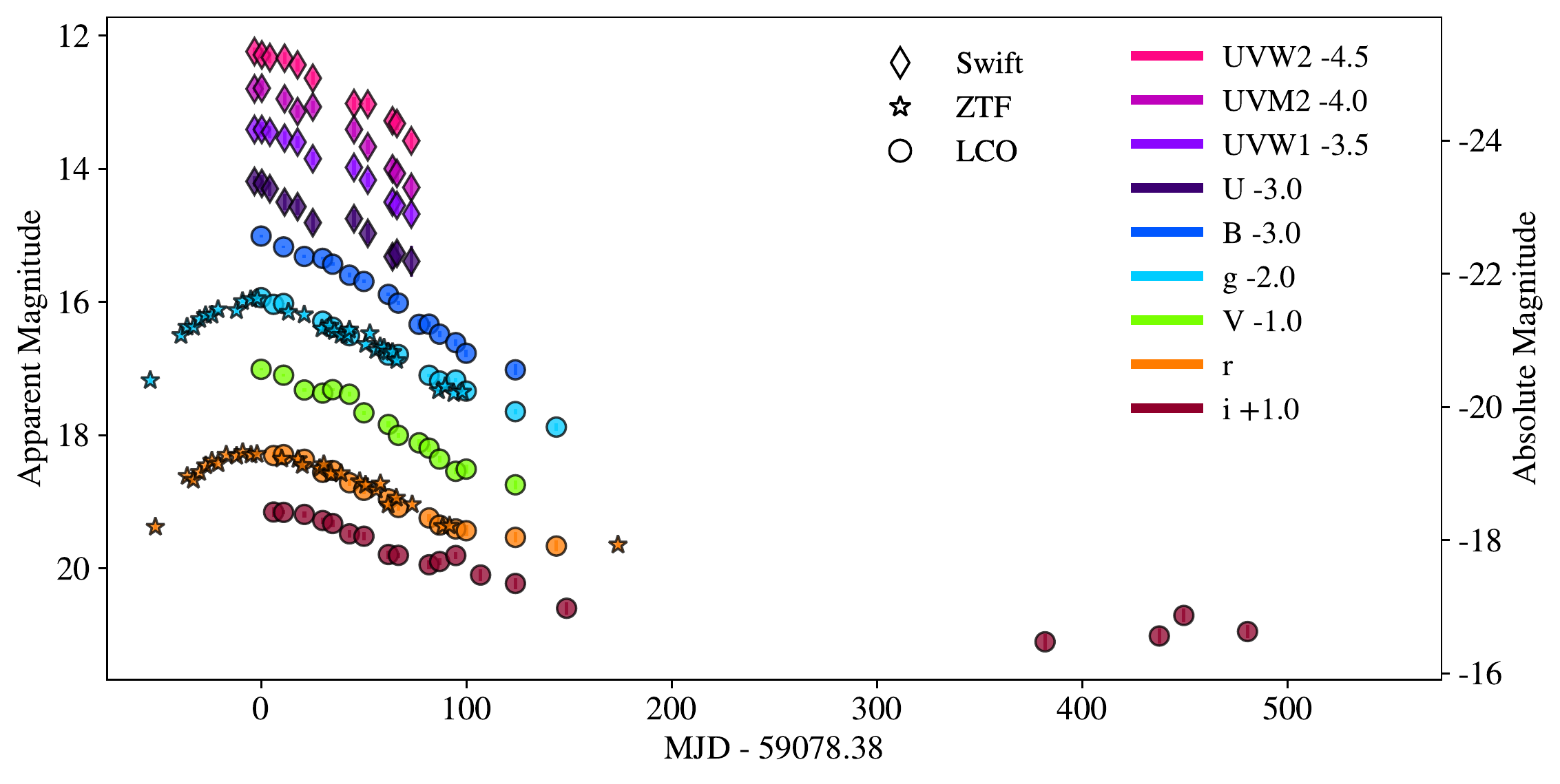}
    \caption{The host-subtracted and Galactic-extinction corrected optical and UV light curves of AT 2020mot as observed by Las Cumbres Observatory, ZTF, and Swift. The light curve in each filter is offset from one another for clarity, and the ZTF $r$-band is further offset by 0.244 mag to account for the systematic difference between the ZTF and Las Cumbres $r$-band filter curves (see Appendix). The $gri$ points from Las Cumbres were subtracted with PS1 archival templates (MJDs 55914.0, 55957.0, and 55991.0, respectively) including the epochs past +200 days from $g$-band peak which show residual TDE flux.}
    \label{fig:offset_light curve_all}
\end{figure*}

\begin{figure*}[t!]
    \centering
    \includegraphics[scale=0.75]{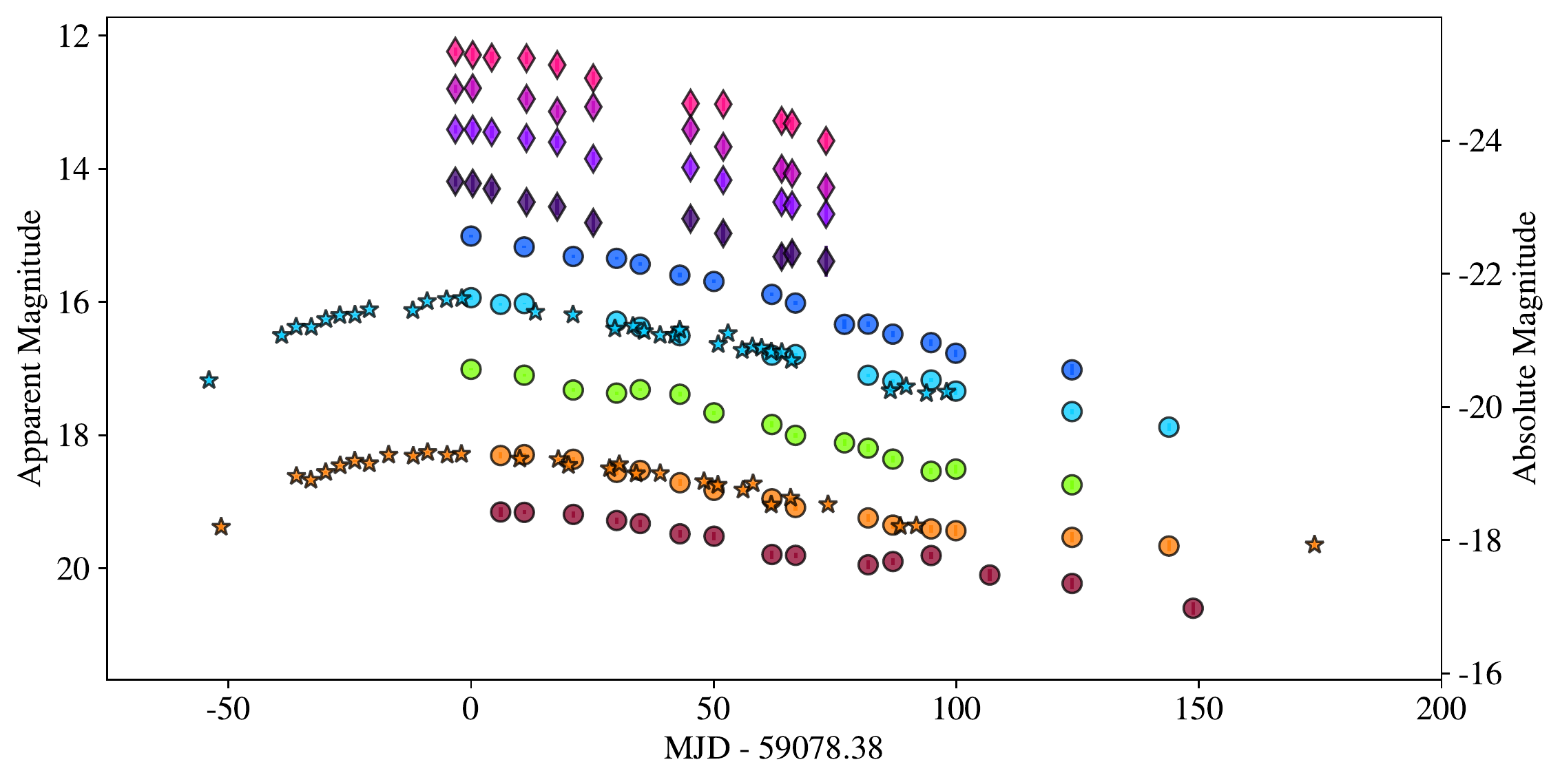}
    \caption{Same as Figure \ref{fig:offset_light curve_all}, excluding the epochs past +200d for clarity of the first optical peak.}
    \label{fig:offset_light curve_early}
\end{figure*}

Host subtraction isolates the evolution of the flare alone without contamination from the host galaxy. Upon subtraction of the science images with Las Cumbres images taken at +490d from $g$-band peak, comparison to the ZTF forced photometry in $gr$-bands at the same epochs, the Las Cumbres subtracted images showed a subtle but clear over-subtraction, indicating residual TDE light. We thus adjusted \texttt{lcogtsnpipe} to use archival images (MJDs 55914.0 in $g$-band, 55957.0 in $r$-band, and 55991.0 in $i$-band) from the Panoramic Survey Telescope and Rapid Response System, Pan-STARRS1 \citep[PS1][]{2010SPIE.7733E..0EK} as templates for image subtraction in the $gri$-bands. We have since acquired several epochs of Las Cumbres data intended for use as templates, most recently at +800 days from $g$-band peak, all of which show residual TDE flux in the $i$-band when subtracted with PS1 templates, while $gr$-bands had entirely faded and showed no residual flux from PS1 subtractions. Indeed, using the +800d images as subtraction templates produce a light curve that is 0.3 mags fainter in $i$-band compared to equivalent subtractions with PS1 templates. One example of the residuals seen in the subtracted images is given in Figure \ref{fig:ps1_residual}. We therefore find that the TDE was still emitting light above baseline host level in the $i$-band at the end of 2022, and use PS1 templates to subtract all $gri$-band data uniformly. We note that subtracting the $gr$-band data with both Las Cumbres and PS1 templates gives the same results, affirming the reliability of the PS1 subtractions.

For $BV$-bands, the +800d images from Las Cumbres were used as templates. All subtraction was performed using HOTPANTS\footnote{\url{https://github.com/acbecker/hotpants}} image subtraction algorithm \citep{1998ApJ...503..325A, 2015ascl.soft04004B}, with normalization to the science image and convolution to the template image.

Supplemental photometry in the $gr$-bands was obtained from the ZTF forced photometry server\footnote{\url{https://ztfweb.ipac.caltech.edu/cgi-bin/requestForcedPhotometry.cgi}} \citep{2019PASP..131a8003M}. These data cover the first detection at MJD = 59014.39 and the subsequent rise to peak, while Las Cumbres data begin near $g$-band peak at MJD = 59078.38. We visually checked the ZTF difference images for non-detections and otherwise excluded data with magnitude errors greater than 0.1.

We have also corrected all photometry for Galactic extinction with $A_V=0.238$ mag \citep[][]{2011ApJ...737..103S} using the Calzetti dust law \citep[][]{Calzetti2000}. Figure \ref{fig:offset_light curve_all} shows the Las Cumbres $BgVri$ and ZTF $gr$ data across all epochs, and Figure \ref{fig:offset_light curve_early} shows the same data until +200 days after optical peak to highlight the bulk of photometry.

There is a systematic difference between the ZTF and Las Cumbres $r$-band magnitudes, due to their filter differences (See Appendix). ZTF $r$-band magnitudes are on average 0.244 mag brighter than Las Cumbres $r$-band data when generating synthetic photometry from spectra of the TDE. The Las Cumbres and ZTF $r$-band subtracted photometry were found to always agree within this range across epochs, but for visual purposes, the ZTF data are shown corrected for this offset to better match Las Cumbres data in figures.

The Las Cumbres data show a bump in the $i$-band (at +83d as seen in \ref{fig:offset_light curve_early}) and late-time emission can be seen, neither of which are typical of TDEs. We expand on the implications of these observations in Section \ref{sec:dust}.

\subsection{Swift Photometry}\label{sec:swift}

UV observations were obtained with the Ultra\-violet Optical Telescope \citep[UVOT][]{2005SSRv..120...95R} from Swift in the UVW2, UVM2, UVW1, and U filters beginning MJD 59075.14 (PI: Gezari), near the time of maximum light. The data were reduced with the Swift Ultraviolet/Optical Super\-nova Archive pipeline \citep{2014Ap&SS.354...89B}, using the aperture corrections and zero-points of \citet{2011AIPC.1358..373B}. Despite low contamination expected from the UV-dim host galaxy, host subtraction was performed with templates taken on March 26 2022 (MJD 59664.0). Swift photometry is presented in Vega magnitudes alongside Las Cumbres and ZTF photometry in Figures \ref{fig:offset_light curve_all} and \ref{fig:offset_light curve_early}. 

\begin{figure}[t]
    \includegraphics[scale=0.5]{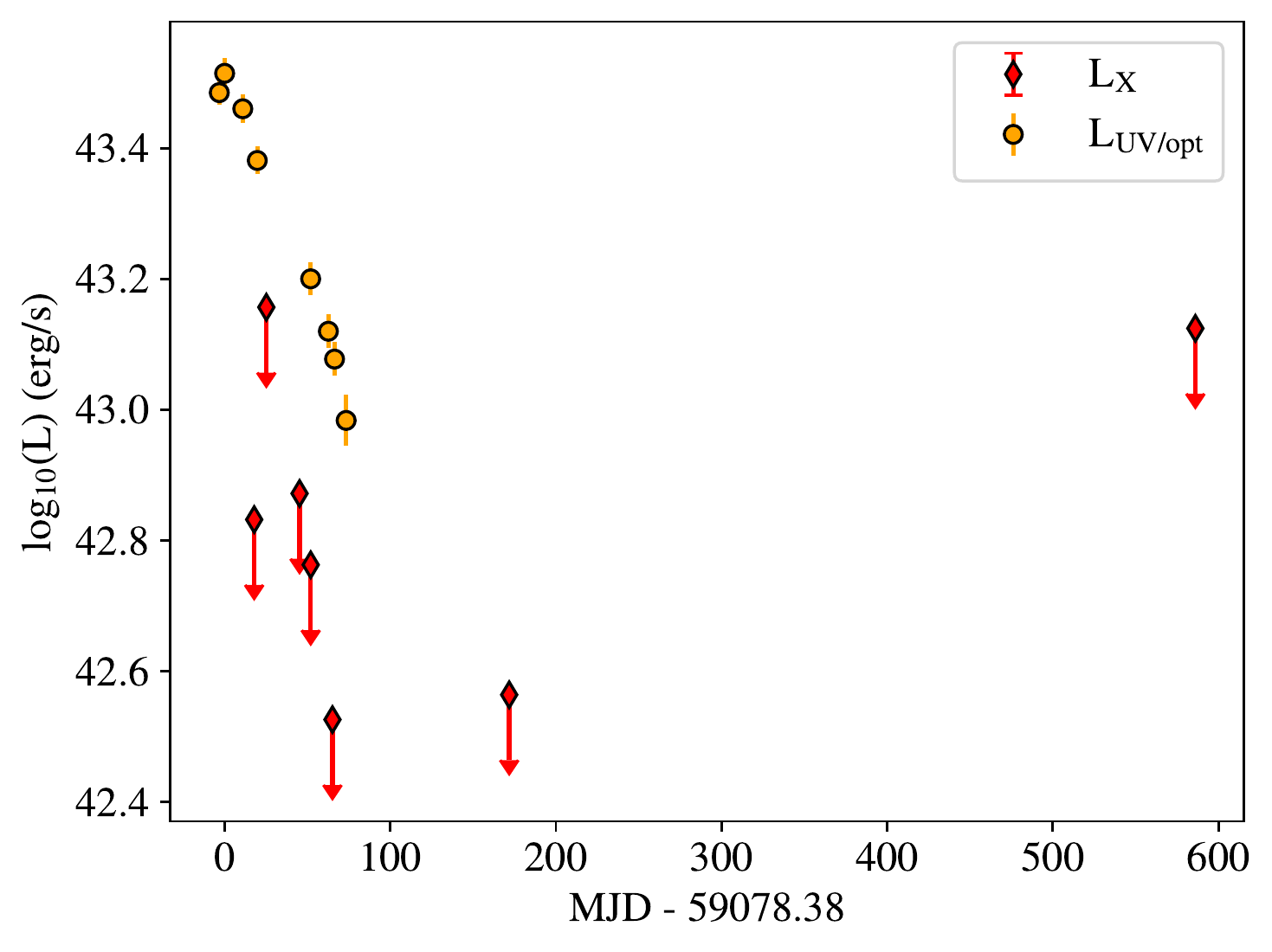}
    \caption{The upper limits of X-ray luminosity from non-detections of AT 2020mot from Swift's XRT (red-filled diamonds), compared with the bolometric luminosity inferred from blackbody fits to the UV/optical light curve (orange-filled circles). The X-ray upper limits show a maximum of $1.33 \times 10^{43}$ erg/s throughout the UV/optical flare.}
    \label{fig:xray_lc}
\end{figure}

Swift's X-ray Telescope (XRT) simultaneously observed AT 2020mot during UVOT follow-up. Following the Swift XRT Data Reduction Guide\footnote{\url{https://www.swift.ac.uk/analysis/xrt/files/xrt_swguide_v1_2.pdf}}, we processed cleaned X-ray event files with \texttt{xselect}\footnote{\url{https://www.swift.ac.uk/analysis/xrt/xselect.php}} to create a light curve binned by an exposure time of 2ks. In this case, zero counts were measured across all observations, so only upper-limits are inferred. We display these upper-limits alongside the bolometric luminosity (see Section \ref{sec:lightcurve}) of AT 2020mot in Figure \ref{fig:xray_lc}. These results are in agreement with the findings of \citealt{Hammerstein2023}.

\subsection{WISE photometry}

The host galaxy of AT 2020mot, WISEA J003113.52 +850031.8, was first observed by WISE in its NEOWISE survey in 2010, and its observations resumed in 2013 when the survey was reactivated as NEOWISE-R \citep[][]{Mainzer2011}. We collected $W1$ ($3.4 \mu$m) and $W2$ ($4.6 \mu$m) data from the Infrared Sci\-ence Archive\footnote{\url{https://irsa.ipac.caltech.edu}}, resulting in 20 epochs of MIR photo\-metry each separated by six months, five of which were taken after the TDE flare.

\begin{figure}[t]
    \centering
    \includegraphics[scale=0.48]{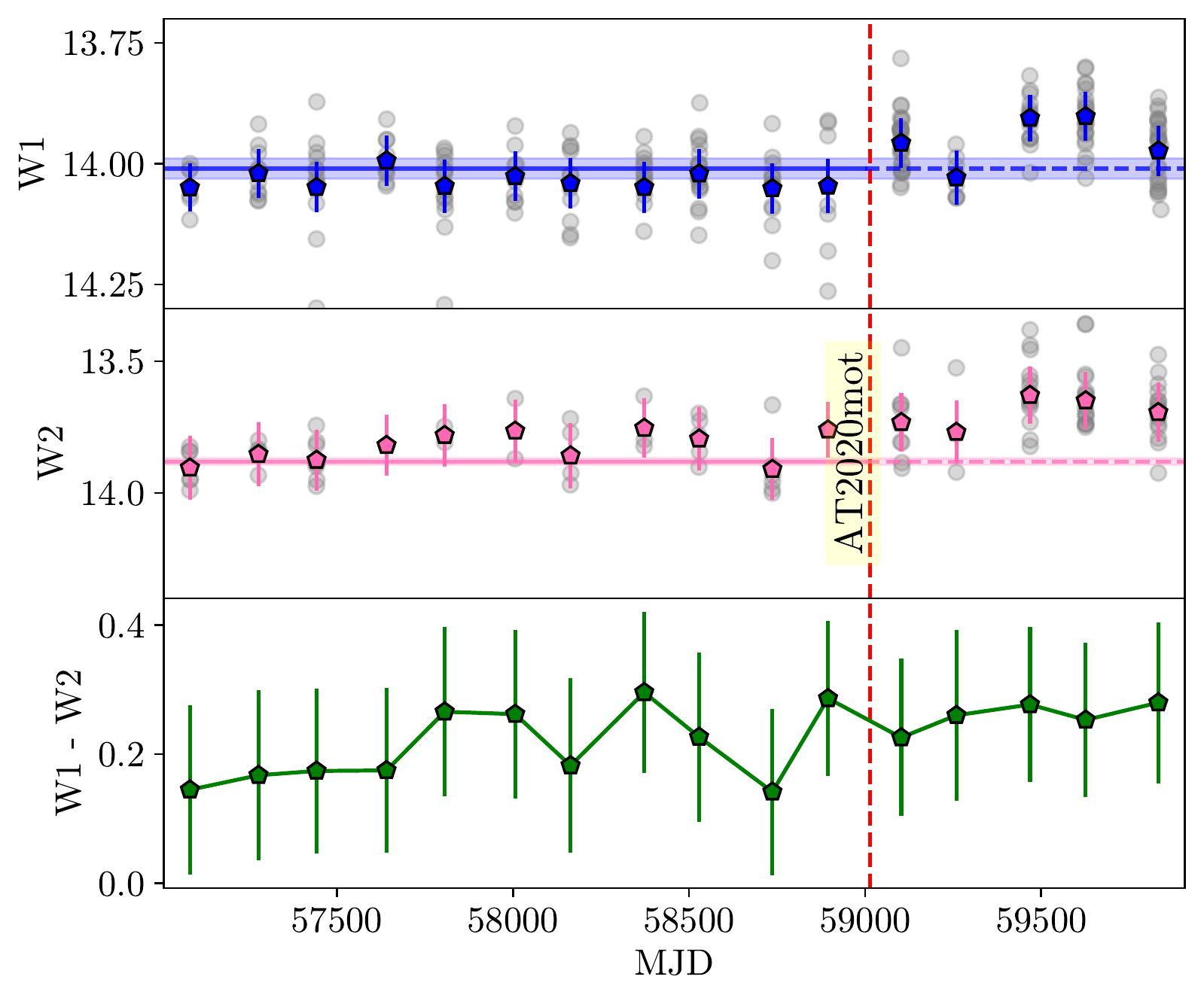}
    \caption{NEOWISE observations at the location of AT 2020mot in $W1$ (top panel) and $W2$ (middle panel) infrared bands. All observations are shown in gray, while the median of each epoch is in blue ($W1$) and pink ($W2$). The vertical dashed red line denotes the time of AT 2020mot first optical detection. The horizontal solid-to-dashed line indicates the median of all quiescent stages in each filter, and the filled region shows the one-sigma range of the quiescent magnitude, clarifying the increase in brightness after the tidal disruption. The bottom panel shows the $W1 - W2$ color, which never reaches AGN levels.}
    \label{fig:WISE_evolution}
\end{figure}

WISE data was parsed for detections with good-quality frames ($\texttt{qi\_fact} > 1$), no contamination and confusion ($\texttt{cc\_flag} = 0$) and with magnitude errors $<0.15$ mag; these requirements filtered out $\sim 10\%$ of observations across all epochs. Figure \ref{fig:WISE_evolution} shows the most recent 10 epochs, including both raw data and binned magnitudes per epoch. We also retrieved time-resolved coadded images of the field created as part of the unWISE project \citep{Lang2014AJ....147..108L, Meisner2018AJ....156...69M} to determine the quiescent host magnitude in both bands before the onset of AT 2020mot.

\begin{figure}[t]
    \centering
    \includegraphics[scale=0.52]{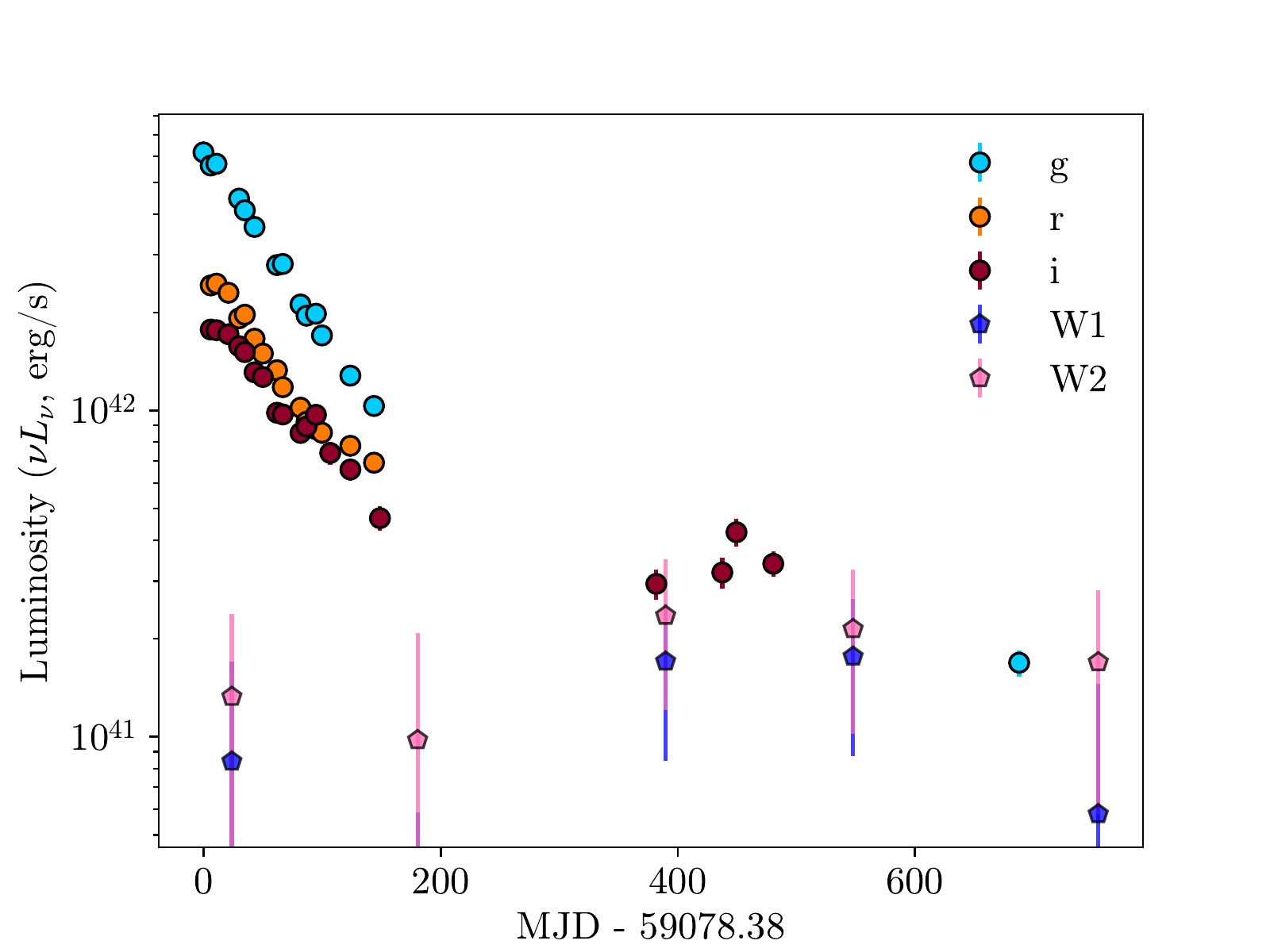}
    \caption{The host-subtracted luminosities of $W1$ (blue pentagons) and $W2$ (pink pentagons) IR-bands, shown alongside Las Cumbres and ZTF $gri$ (cyan, orange, and red circles respectively) subtracted photometry for reference relative to g-band peak. Only one epoch in each band of WISE photometry, before 2022, shows significant excess flux compared to the pre-TDE quiescent stage (the second pre-2022 epoch is within the errors of the quiescent stage).}
   \label{fig:WISE_subtracted}
\end{figure}

No significant variability is detected between each epoch before MJD 59000, hence we consider all data before the detection of AT 2020mot to be the quiescent stage. We display each epoch's median magnitudes alongside the magnitude of the quiescent stage in Figure \ref{fig:WISE_evolution}, which is measured from taking the median of all epochs in the corresponding filter preceding the TDE detection. We also present the $W1 - W2$ color, which never exceeds the AGN color-cut of $W1 - W2 \geq 0.8$ \citep[][]{Stern2012}.

For the post-discovery epochs, we subtract the contribution from the host galaxy to the WISE measurements during the epochs with TDE contribution. However, as can be seen before subtraction in Figure \ref{fig:WISE_evolution}, the first two epochs after TDE detection, in both $W1$ and $W2$ filters, are still within one sigma of the quiescent stage. We thus only consider three epochs as detection of enhanced infrared brightness after the TDE occurred by exceeding the quiescent flux by 1.5 sigma. This final epoch, subtracted with the quiescent flux, gives integrated luminosities across the respective bands $L_{W1} = 2.09 \pm 0.87 \times 10^{41}$ erg/s and $L_{W2} = 1.83 \pm 1.22 \times 10^{41}$ erg/s. The subtracted $W1$ and $W2$ luminosities are displayed in Figure \ref{fig:WISE_subtracted}, with $gri$-band photometry from Las Cumbres shown for comparison.

\subsection{Spectroscopy}\label{subsec:spec}

\begin{figure}[t!]
    \centering
    \includegraphics[scale=0.57]{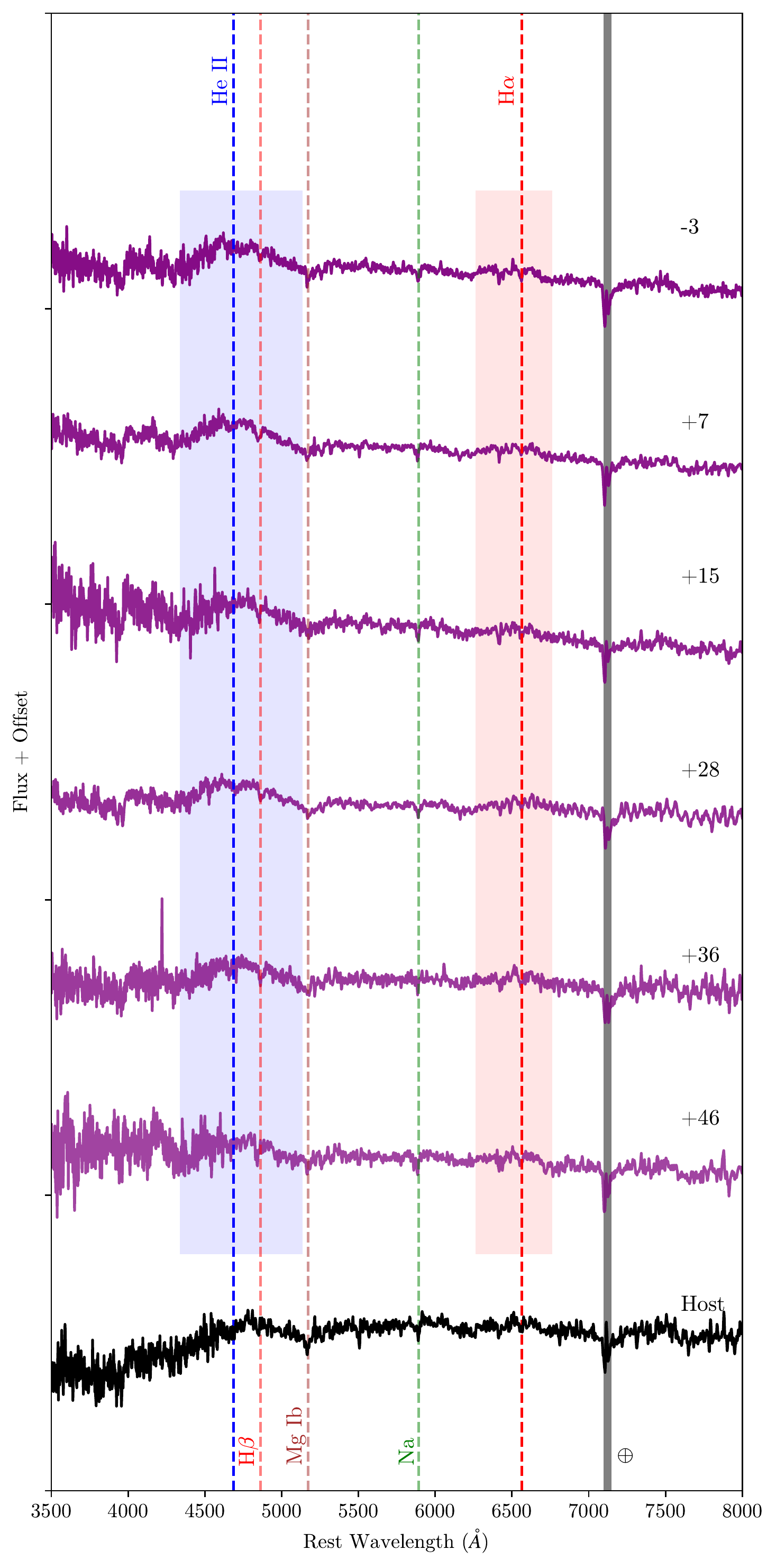}
    \caption{AT2020mot and host spectra from Las Cumbres' FLOYDS-FTN. All spectra have been corrected for Galactic extinction. Six epochs during the tidal disruption (purple lines) range from 3 days before $g$-band peak to 46 days after, offset for visual clarity, and a host spectrum (obtained 726 days after peak) is shown below (black line). The vertical dashed lines mark characteristic lines from the TDE and host, with H$\alpha$ and H$\beta$ in red, He II $\lambda$4686 $\angstrom$ in blue, and host Na $\lambda$5890 $\angstrom$ and Mg Ib $\lambda$5167, $\lambda$5172, $\lambda$5183 $\angstrom$ triplet in green and brown, respectively. The shaded regions indicate the breadth of H$\alpha$ and He II. We also label the telluric contamination.}
    \label{fig:spectra}
\end{figure}

\begin{figure}[t!]
    \centering
    \includegraphics[scale=0.57]{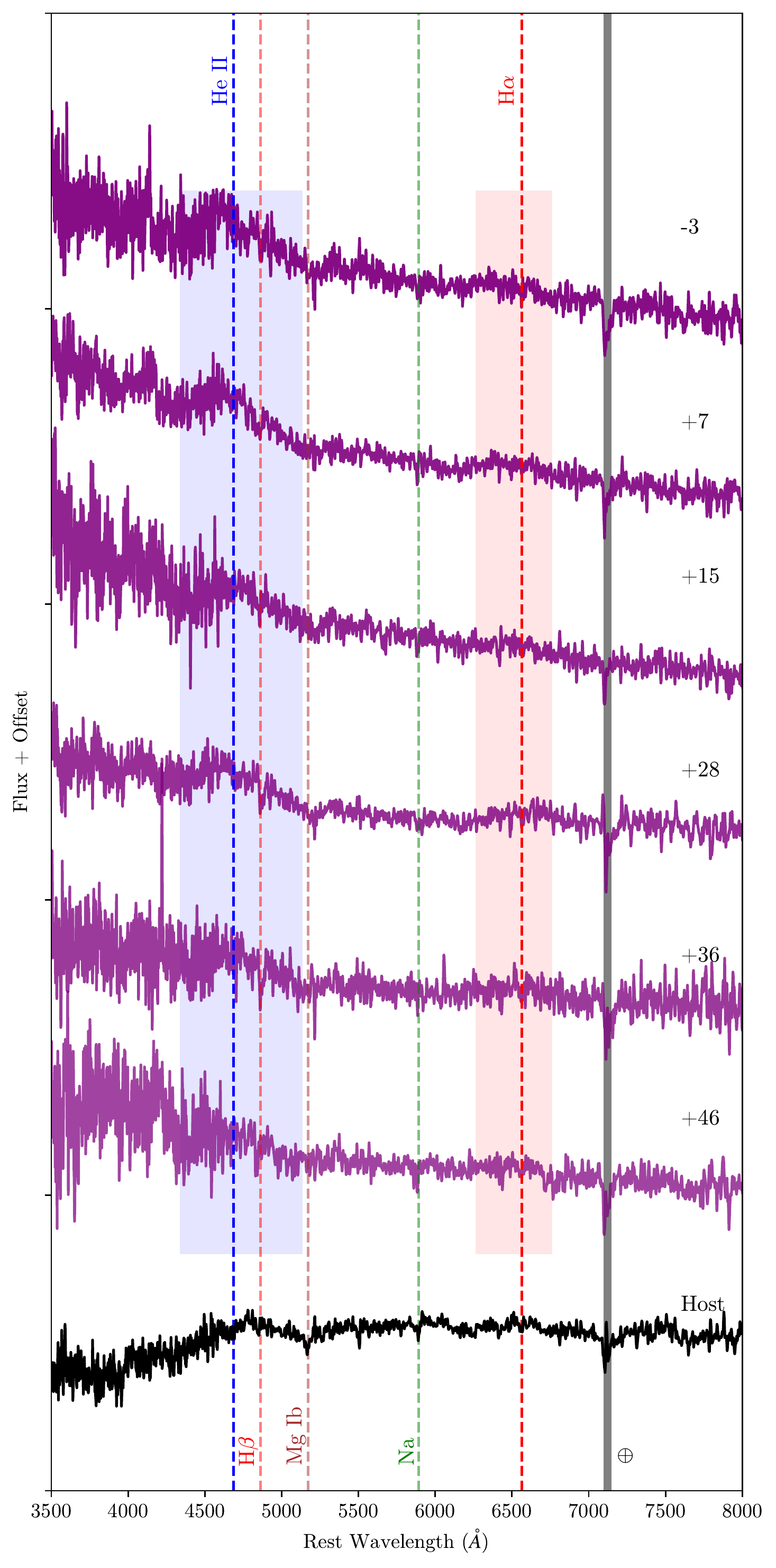}
    \caption{Same as Figure \ref{fig:spectra}, but each epoch of AT 2020mot (purple) has been host-subtracted as described in Section \ref{subsec:spec}. This process reveals the blue continuum as observed photometrically, typical for TDEs, and further distinguishes the transient features from the host contamination. This calibrating subtraction also highlights the broad He II $\lambda$4686 $\angstrom$ that is present from the TDE and fades by 46 days after peak. However, the subtraction exaggerates noise, especially on the wavelength edges due to red fringing and lower blue resolution, so any features blueward of He II are indistinguishable.}
    \label{fig:spectra_warped}
\end{figure}

\begin{figure}[t!]
    \centering
    \includegraphics[scale=0.4]{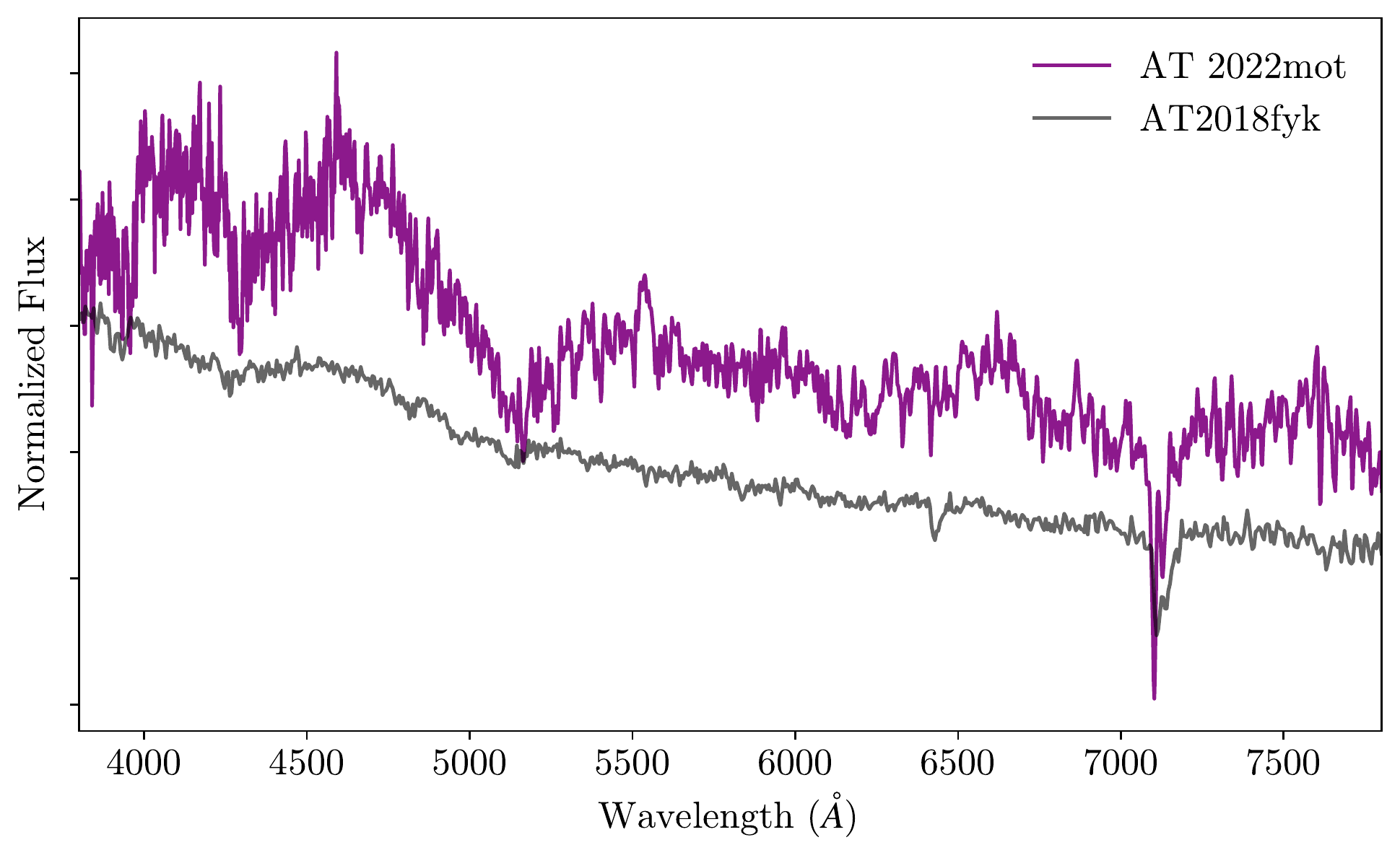}
    \caption{AT2020mot's host-subtracted spectrum from 7 days after peak (purple) compared to the AT2018fyk classification spectrum (gray). Both are normalized such that their integrated flux equals 1 for visual comparison. AT2018fyk is another TDE with broad He II $\lambda$4686 \citep{2018TNSCR1397....1W, 2019MNRAS.488.4816W}}.
    \label{fig:speccomparison_18fyk}
\end{figure}

Optical spectra were taken with FLOYDS on Las Cumbres’s 2m Faulkes Telescope North (FTN) on Haleakal$\bar{a}$, HI, reduced using the \texttt{floydsspec} pipeline\footnote{\url{https://github.com/svalenti/FLOYDS_pipeline/}}, as described in \cite{2014MNRAS.438L.101V}. The pipeline performs flux and wavelength calibration, cosmic-ray removal, and final spectrum extraction. The original spectra cover 3500 to 10,000 $\angstrom$, at a resolution $R \approx 300-600$ and epochs from -3 to +46 days with respect to the $g$-band peak at MJD = 59078.38, calculated by fitting a template TDE light curve to the observations with a least-squares method and finding the time of maximum light from the template. We subsequently obtained a host spectrum at +726 days after peak. The dates and phases of each spectrum we report are listed in Table \ref{table:specdates}

Each spectrum was taken at airmass $> 2.3$, compared to airmass $1.8-2.0$ for photometry taken at higher-latitude sites. We present spectra without host subtraction in Figure \ref{fig:spectra}, and with the host spectrum, taken 726 days after the peak of AT 2020mot, presented alongside for comparison.


In Figure \ref{fig:spectra_warped}, we also show the six epochs of AT 2020mot spectra after host-subtraction via calibration to match the host-subtracted $BgVr$ photometry. We first subtracted the host spectrum from each transient epoch multiplied by a factor that minimizes the difference between the synthetic photometry and the real host-subtracted photometry of the corresponding epoch using the sum of all bands. We then divided between this host-subtracted synthetic photometry and the observed photometry, resulting in scale factors at the central wavelength of each filter. We interpolated between these filter-specific scale factors using a linear spline to give scale factors across all wavelengths. We multiplied the spectrum by these interpolated scale factors to get the final calibrated spectrum, again minimizing the difference between synthetic and observed photometry at each iteration by using \texttt{scipy.optimize.minimize}. Final calibrated spectra are shown in Figure \ref{fig:spectra_warped}. Finally, we present a comparison of AT2020mot to the classification spectrum of AT2018fyk in Figure \ref{fig:speccomparison_18fyk}. AT2018fyk is a TDE with broad He II $\lambda$4686\citep{2018TNSCR1397....1W, 2019MNRAS.488.4816W}, thus similar in classification to AT2020mot. 

\begin{deluxetable}{cccc}[t!]
\label{table:specdates}
    \tablehead{
    \colhead{MJD} & \colhead{Phase} & \colhead{Airmass} & \colhead{Exposure Time (s)}
    }
    \startdata
        59074 & -3 & 2.31 & 3600\\
        59084 & +7 & 2.31 & 3600\\
        59092 & +15 & 2.32 & 3600\\
        59105 & +28 & 2.32 & 3600\\
        59117 & +36 & 2.32 & 3600\\
        59119 & +46 & 2.31 & 3600\\
    \enddata
    \caption{The MJD and phase at which each reported spectrum was taken by FLOYDS-FTN, where phase is the number of days from peak in $g$-band (at MJD 59078.38).}
\end{deluxetable}

\section{Observational Results}

We assess the multi-band photometric data by comparing the light curve's near-infrared peculiarities against a broad sample of optical TDEs, calculating blackbody radii and temperatures of the event across epochs, and modeling the light curve with two TDE emission mechanisms to determine the masses of the central BH and the stellar progenitor.

\subsection{Near-Infrared Excess and Bump}\label{subsec:bump}

The $i$-band data is brighter than expected across all epochs and shows a ``bump" starting at MJD = 59160 ($+83$ days after peak). The peak of this bump is approximately 100 days after maximum light in the $g$-band. 


\begin{figure}[t]
    \centering
    \includegraphics[scale=0.375]{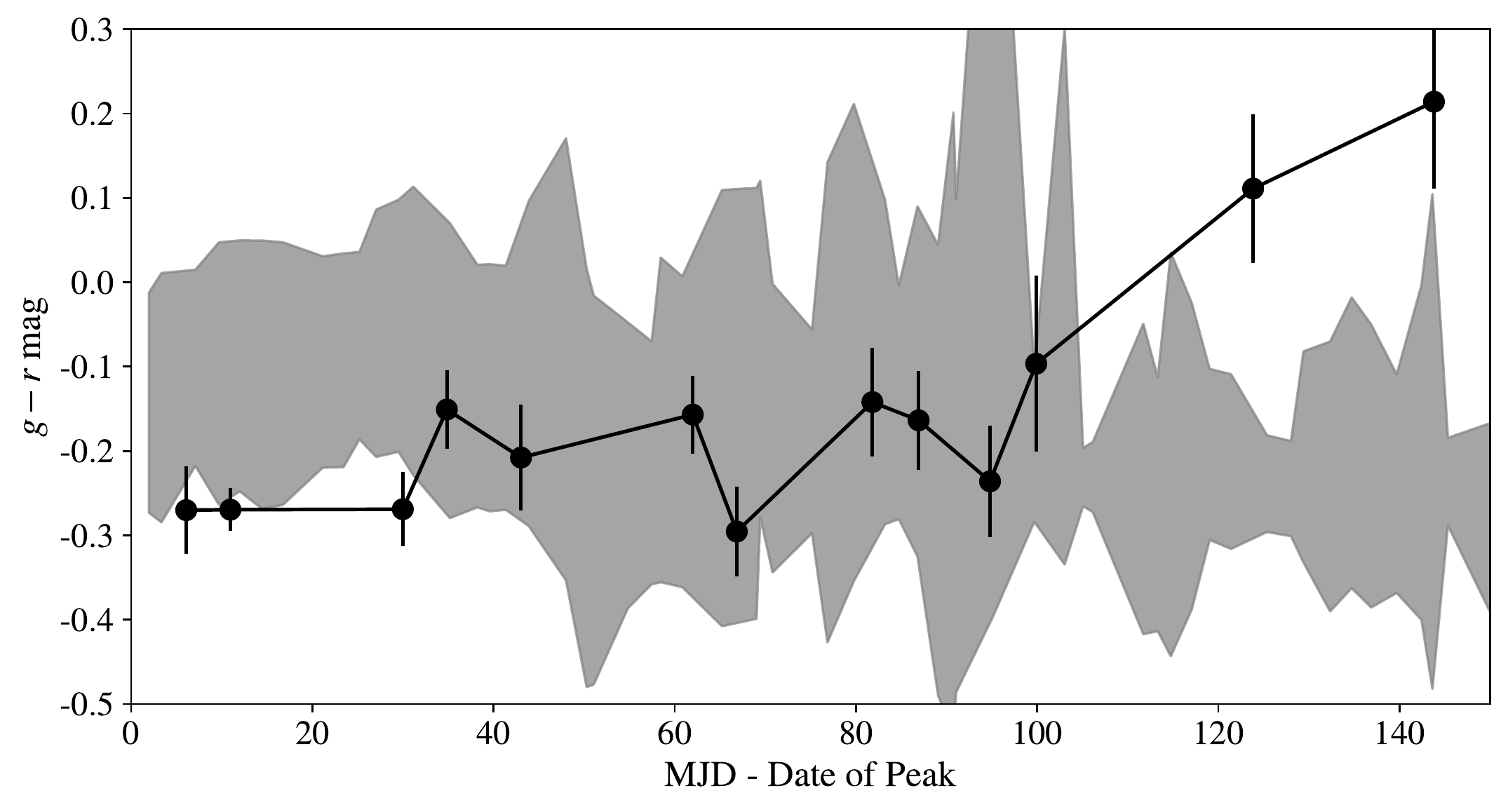}
    \includegraphics[scale=0.375]{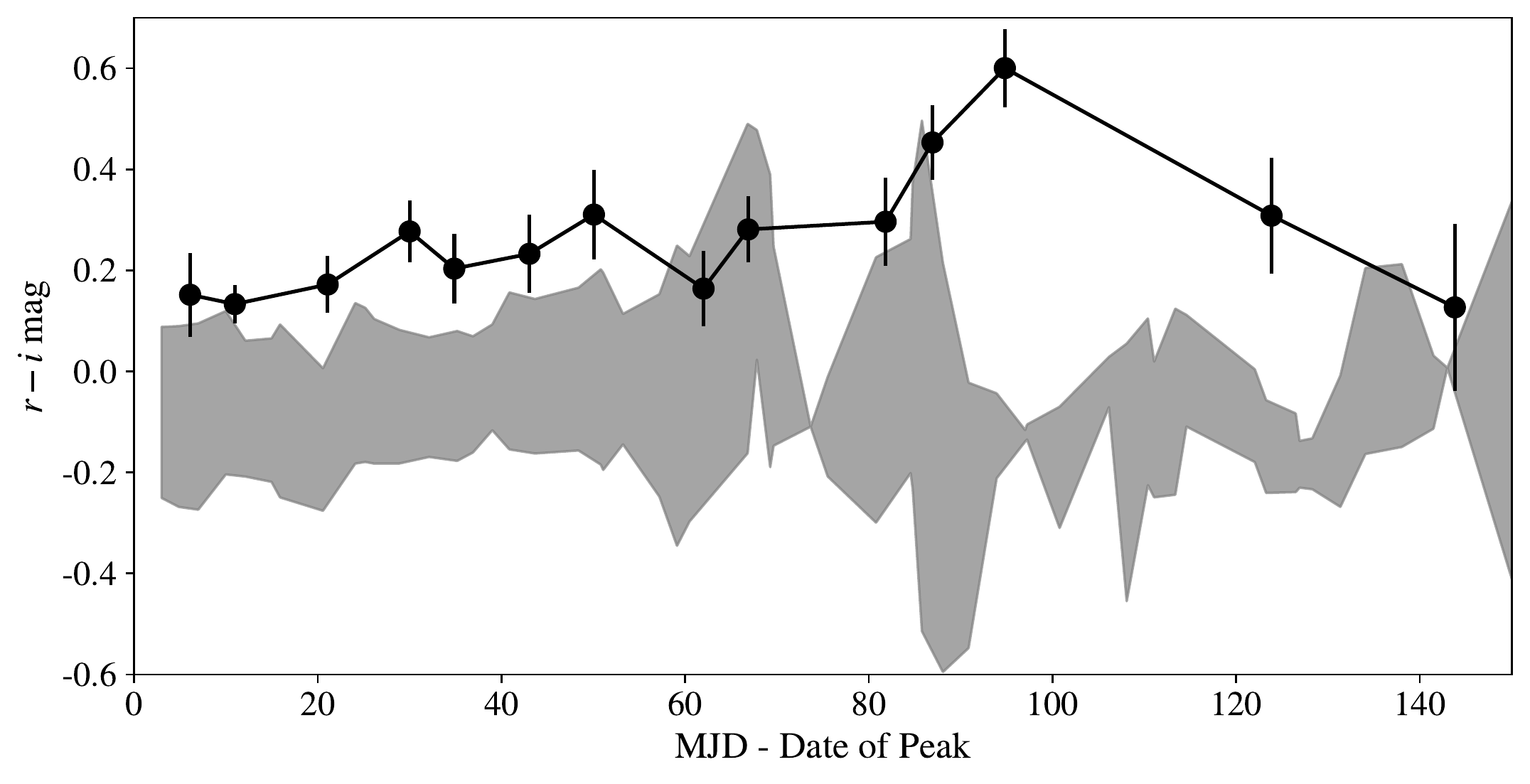}
    \caption{Progression of $g-r$ (top panel) and $r-i$ (bottom panel) colors of AT2020mot (black points) and a sample of 15 archival TDEs (gray filled-in section). Dates are shown with respect to each TDE's $g$-band peak. The archival TDE spread in gray shows the moving median and absolute median deviation of the color among all TDEs at each epoch. AT2020mot's $g-r$ color is unremarkable until lat times, while in $r-i$ is uniquely red across all epochs and reddens even further at the bump +80 days after peak before declining.}
    \label{fig:colors}
\end{figure}

Color evolution illustrates AT\,2020mot's unique red excess, as well. In Figure \ref{fig:colors} we present AT 2020mot's $g-r$ and $r-i$ colors at each epoch, against a backdrop of the respective median colors of 15 archival TDEs with $gri$ photometry for 80 days after first detection: AT2017eqx, AT2019qiz, \citep[][]{Nicholl2019ehx}; AT2018hyz \citep[][]{Gomez2020_18hyz}; AT2018hco, AT2019bhf, AT2019cho, AT2019dsg, AT2019ehz, AT2019meg \citep{VV2021b}; AT2018iih \citep{2020MNRAS.497.1925G}; PS1-10jh \citep[][]{Gezari2012, Gezari2015_10jh}; PS1-11af \citep[][]{Chornock2014}; PTF09ge \citep[][]{Arcavi2014}; iPTF16axa \citep[][]{Hung2017_iPTF16axa}; iPTF16fnl \citep[][]{Blagorodnova2017}. Las Cumbres is the only telescope contributing $i$-band photometry for AT\,2020mot. While AT\,2020mot progresses overall in agreement with the background sample in $g-r$ color without unusual evolution, it has high $r-i$ color compared to the sample throughout the entire light curve decline, as well as increasing to the reddest colors at the time of the bump.

\subsection{Light Curve Fitting} \label{sec:lightcurve}

We fit the UV and optical spectral energy distribution (SED) of AT\,2020mot to a blackbody across its light curve to estimate the evolution of the photospheric radius and temperature. We exclude the $i$-band data from these fits as we interpret their excess brightness and late-time bump as an indication of contribution from a dust echo (see Section \ref{sec:dust}). 

We fit a blackbody spectrum to epochs across the light curve with five or more filters of fitted data per epoch range (2 days), starting from MJD = 59075 near maximum light, using the \texttt{lightcurve\_fitting} package from \cite{griffin_hosseinzadeh_2022_6519623}. The requirement of five filters ensures that only epochs with UV data from Swift will be used to avoid underestimating the temperature (see \citealt{Arcavi2022}). The code uses  Markov Chain Monte–Carlo (\texttt{MCMC}) sampling with the \texttt{emcee} package \citep{2013PASP..125..306F} with a broad log-flat prior of $10,000$ K $\leq T \leq 100,000$ K and $10 R_{\odot}$ $(\sim10^{-7} \text{pc}) \leq R \leq 10^{6} R_{\odot}$ $(\sim 10^{-2} \text{pc})$.

\begin{figure}[t]
    \centering
    \includegraphics[scale=0.5]{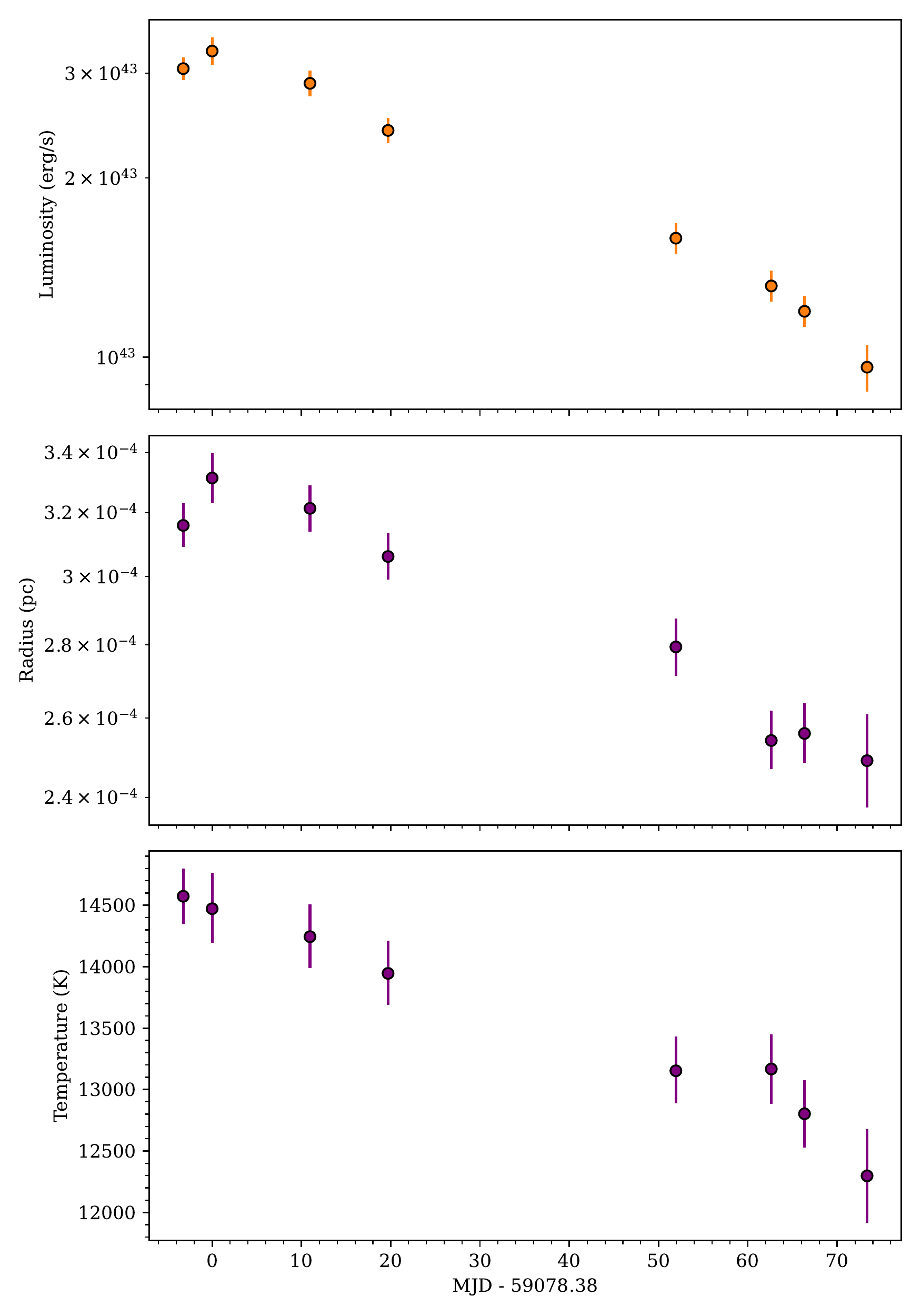}
    \caption{The blackbody temperature (bottom) and radius (middle) as fit from the photometric evolution of AT2020mot using Swift $UVW2, UVM2, UVW1$ and Las Cumbres/ZTF $BgVr$ data. The blackbody luminosity is calculated from the fitted parameters as $L = 4\pi R^{2} \sigma T^{4}$ and plotted in the top panel.}
    \label{fig:blackbody}
\end{figure}

Though this package provides 16-84 percentiles as the errors on estimated parameters (which are plotted for reference in Figure \ref{fig:blackbody}), we quantify our temperature measurements with the uncertainties found in \cite{Arcavi2022} for the bounds on light curve fits to blackbodies with UV and optical data. Thus we find the temperature declines slowly from $T_{\text{BB}} = 14600\pm2500$ K to $T_{\text{\text{BB}}} = 12300\pm2500$ K. These temperatures are consistent with, though on the lower end of, other TDE observations \citep[10000-50000 K; e.g.][]{VanVelzen21}. The blackbody radius is simultaneously found to decrease more strongly, from $R_{\text{BB}} = 3.16\pm0.06\times 10^{-4}$ pc to $R_{\text{BB}} = 2.49\pm0.08\times 10^{-4}$ pc., also consistent with TDE observations \citep{VanVelzen21}. This radius is a factor of 1000 larger than the Schwarzschild radius of $\sim 2\times 10^{-7}$ pc for a SMBH of mass $\sim 10^{6} M_{\odot}$ (see Section \ref{subsec:mosfit}). These parameters give an estimated blackbody luminosity $L_{\text{BB}} = 3.05 \pm 0.20 \times 10^{43}$ erg/s at peak, corresponding to an initial Eddington ratio of $\sim0.1$ for the average SMBH mass inferred in Section \ref{subsec:mosfit}, $M_{BH} \approx 4.5 \times 10^{6} M_{\odot}$, and declining to $L_{\text{BB}} = 9.63 \pm 0.83 \times 10^{42}$ erg/s at late times.

\begin{figure*}[t]
    \includegraphics[scale=0.3]{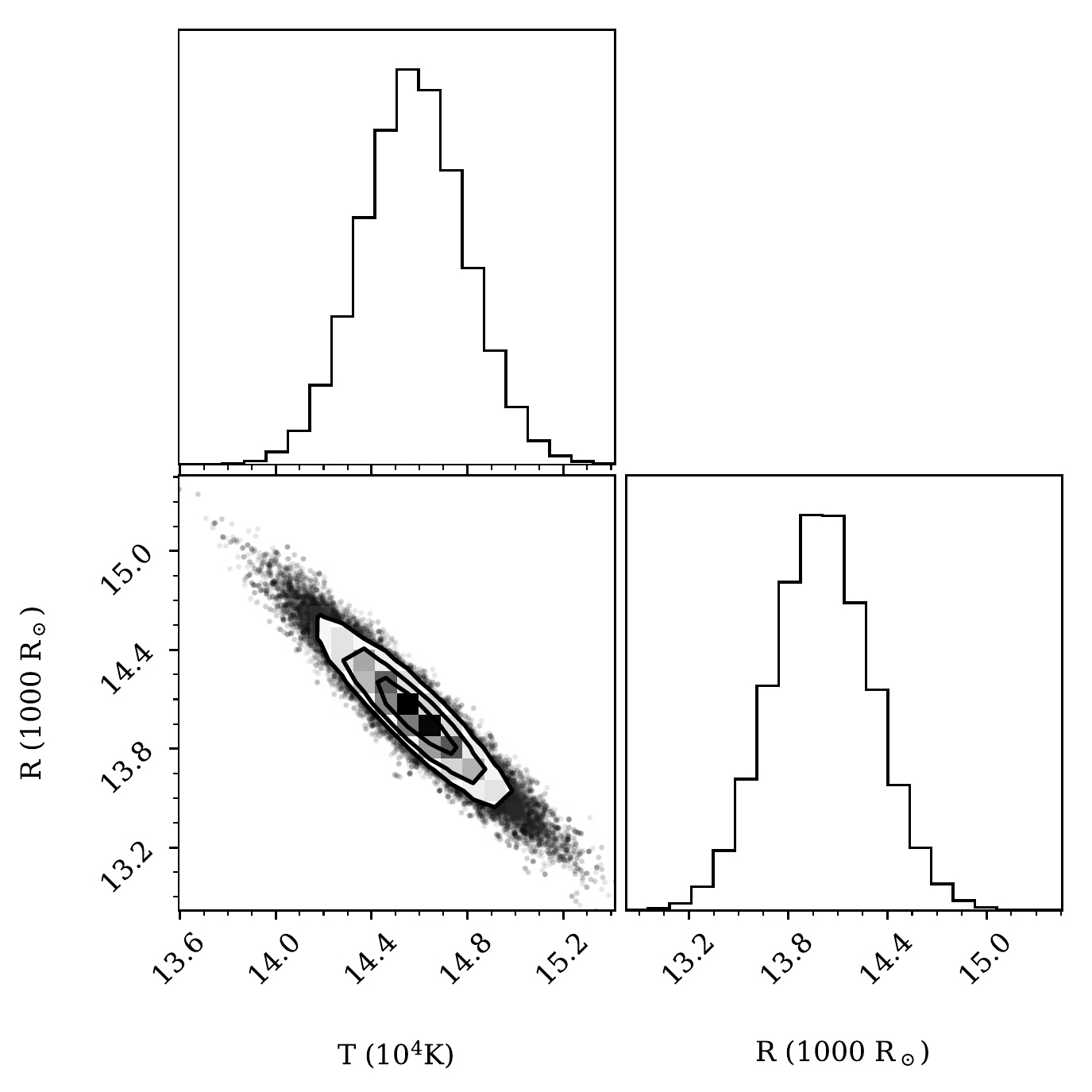}
    \includegraphics[scale=0.3]{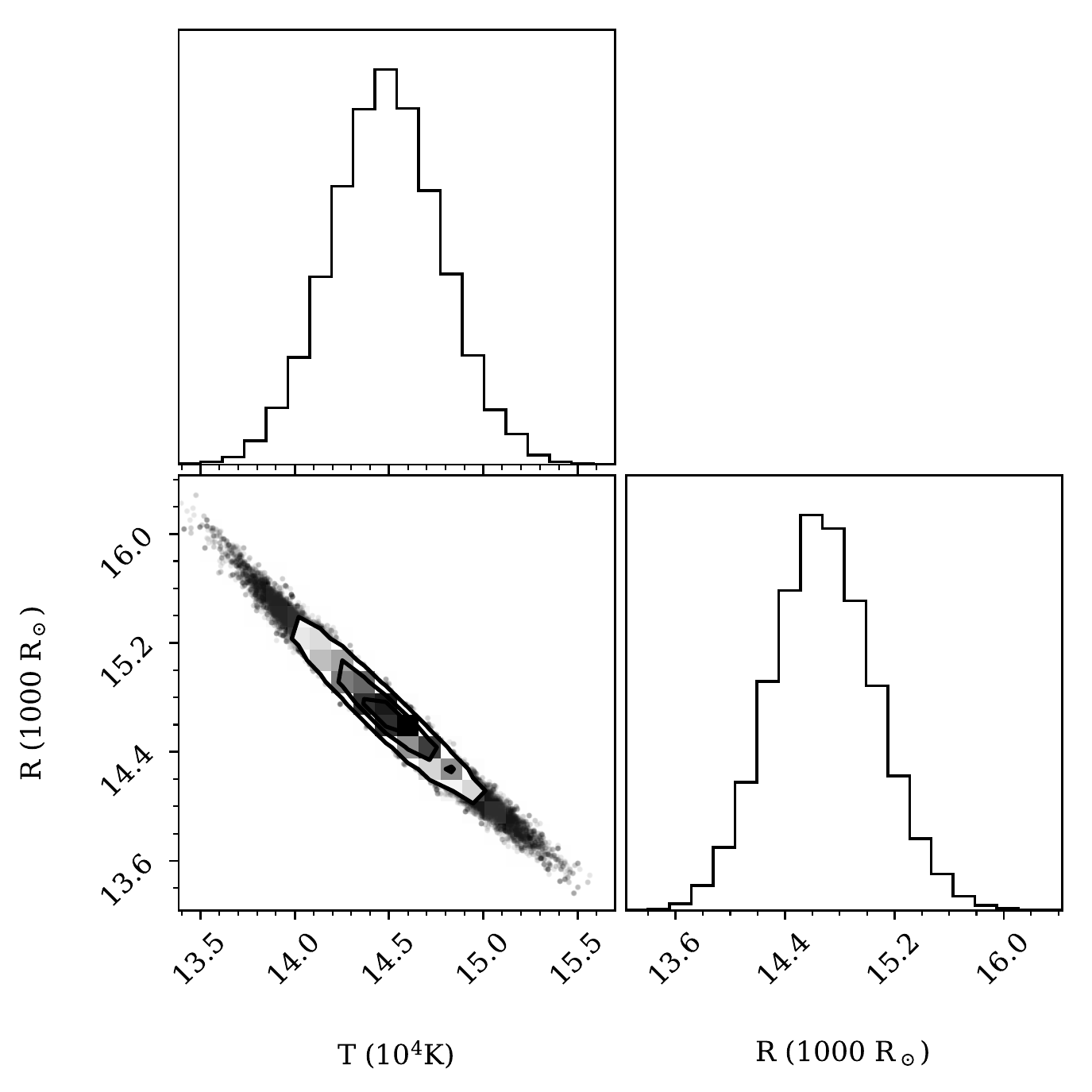}
    \includegraphics[scale=0.3]{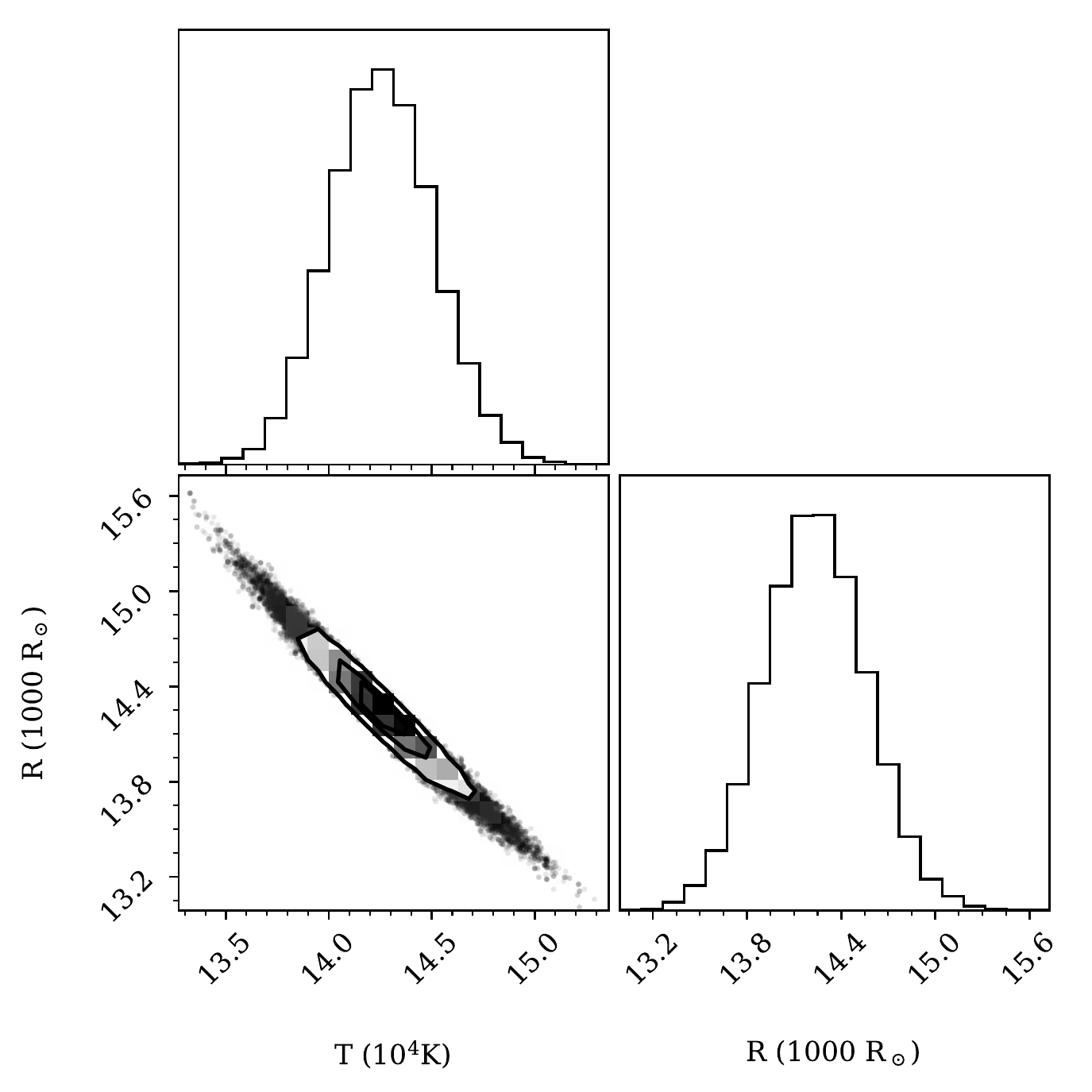}
    \includegraphics[scale=0.3]{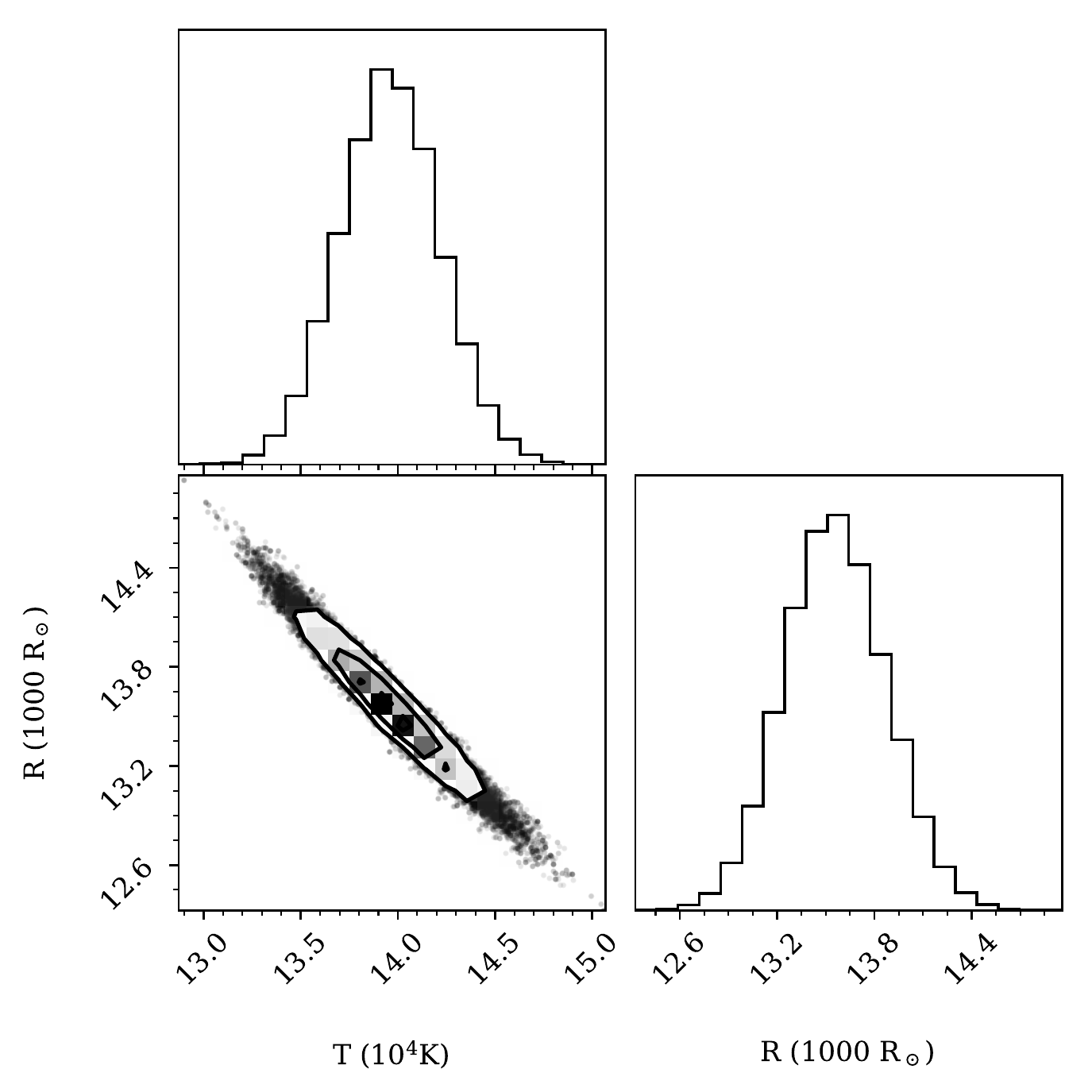}\\
    \includegraphics[scale=0.3]{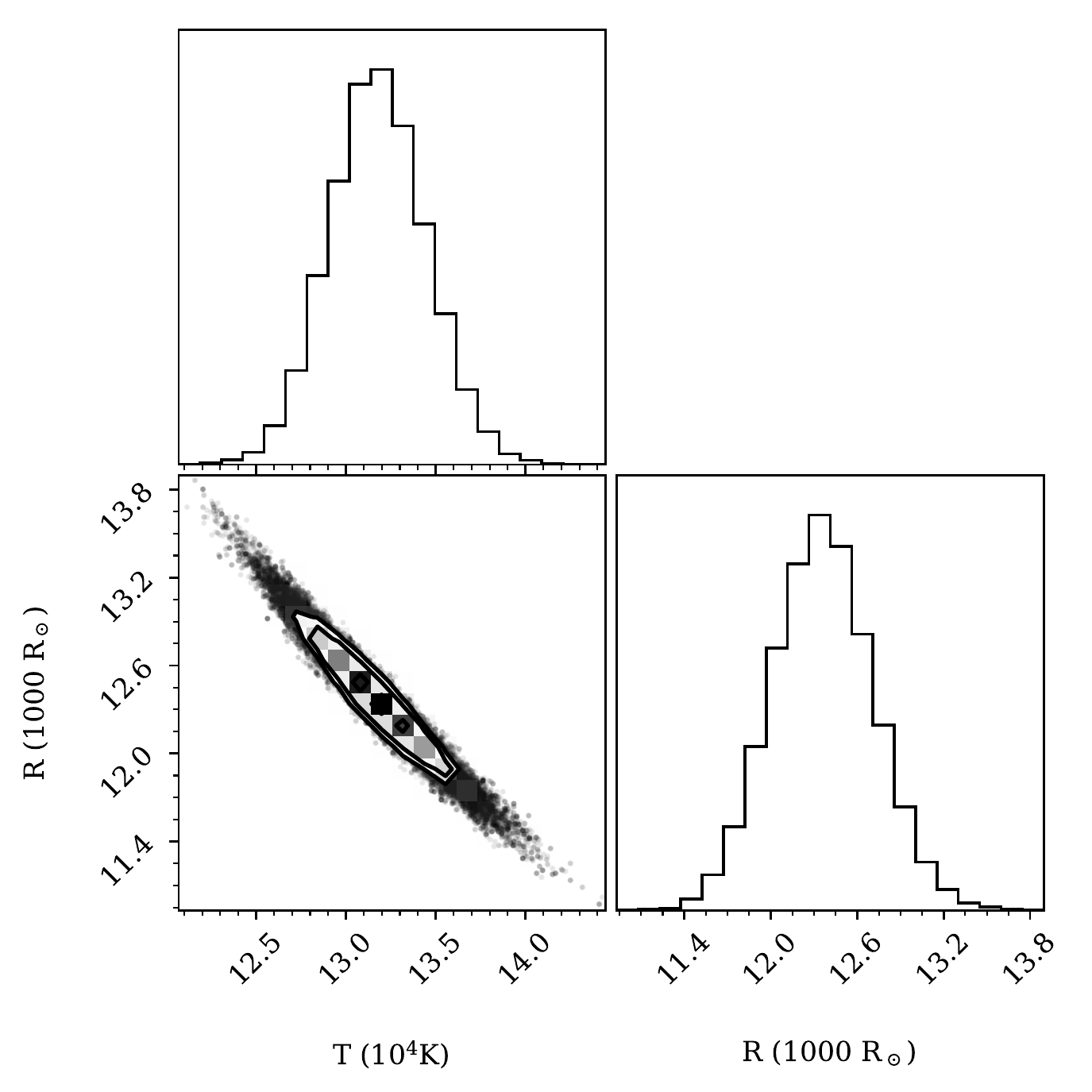}
    \includegraphics[scale=0.3]{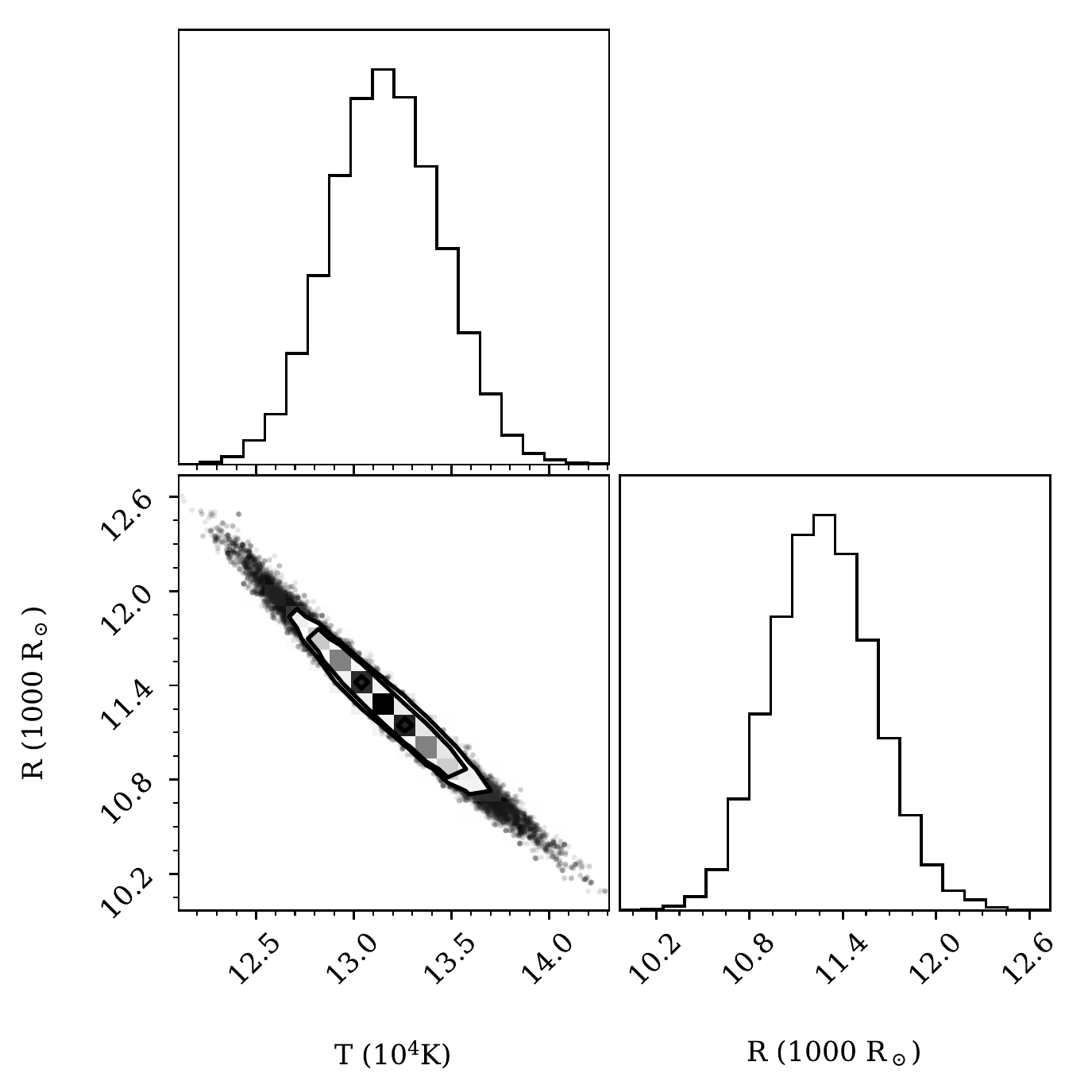}
    \includegraphics[scale=0.3]{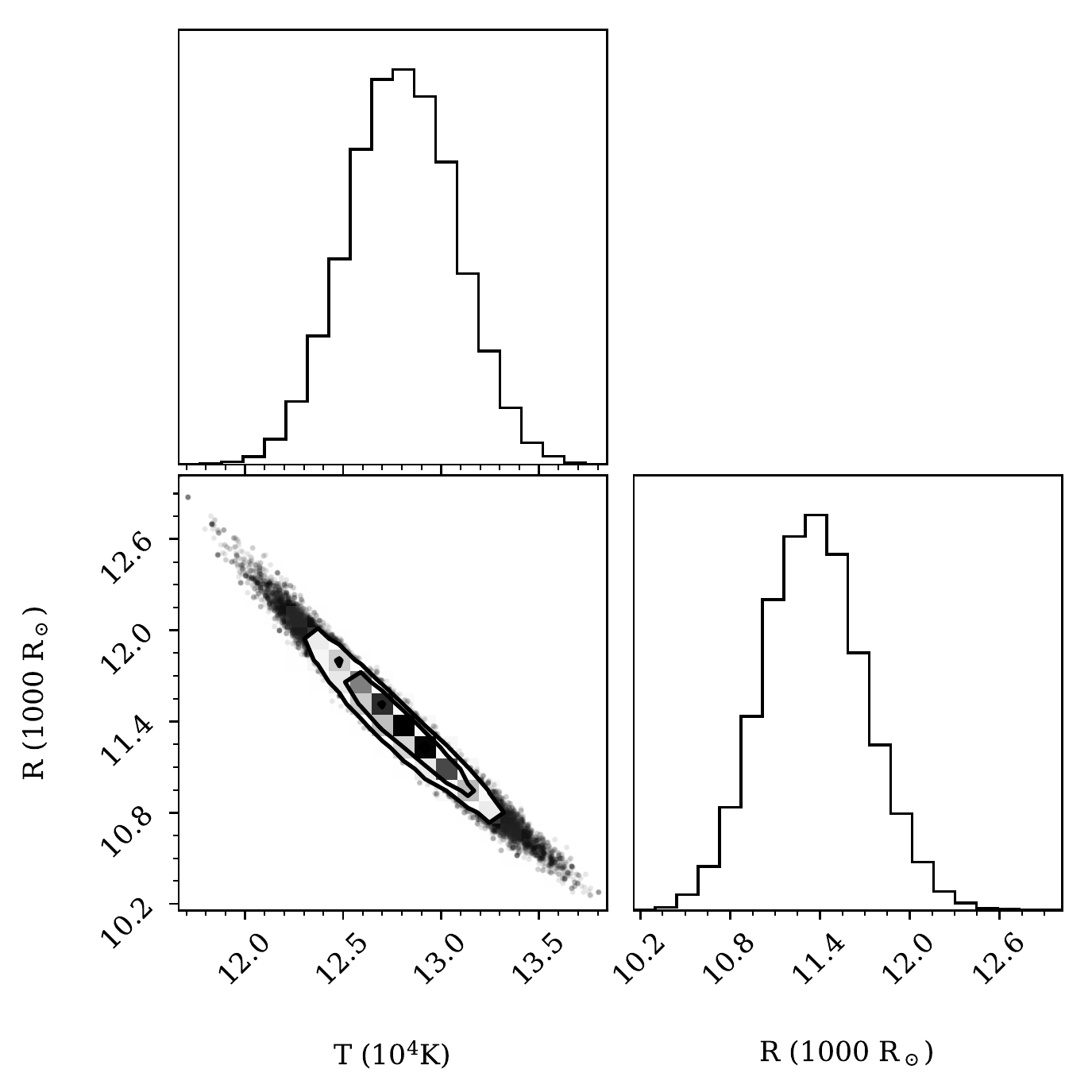}
    \includegraphics[scale=0.3]{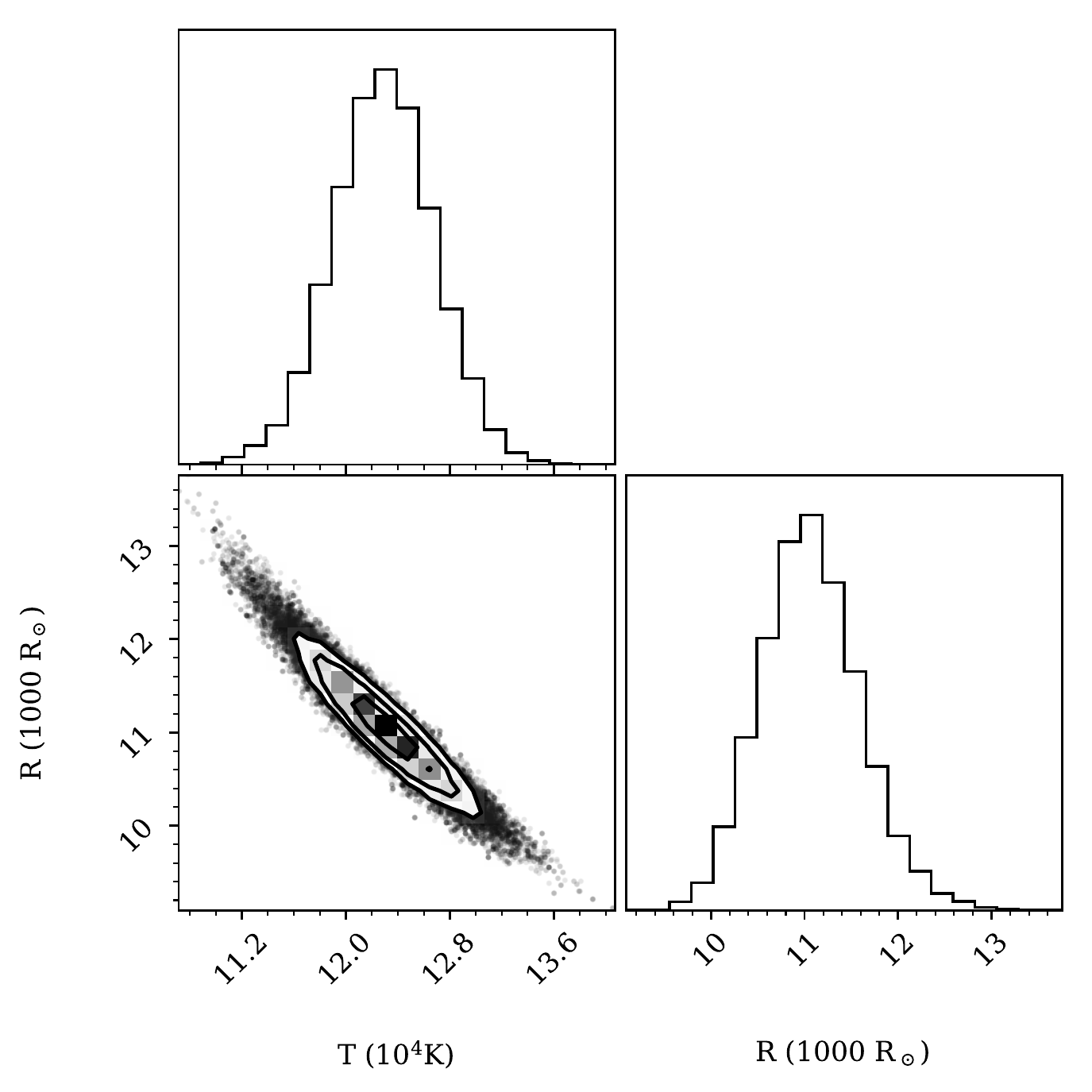}
    \caption{Corner plots of the light curve fit to a blackbody temperature (y-axis, in $10^{4} K$) and radius (x-axis, in $1000 \times R_{\odot}$) to each epoch with 5 filters of data available within a 4-day range. Top, left to right: MJD = 59074.4, MJD = 59078.102, MJD = 59089.47, MJD = 59097.68; bottom, left to right: MJD = 59129.58, MJD = 59131.31, MJD = 59144.80, MJD = 59152.87.}
    \label{fig:blackbody_corner}
\end{figure*}

We cannot discern whether reprocessed accretion emission or stream shocks fuel the flare of AT\,2020mot from our blackbody estimates of the radius and temperature evolution. The lower temperatures we observe could be due to higher extinction in this TDE compared to others, which would be expected in the case of a dust echo.

\subsection{Black Hole and Pre-Disruption Stellar Masses} \label{subsec:mosfit}

\begin{figure*}[t]
    \includegraphics[scale=0.3]{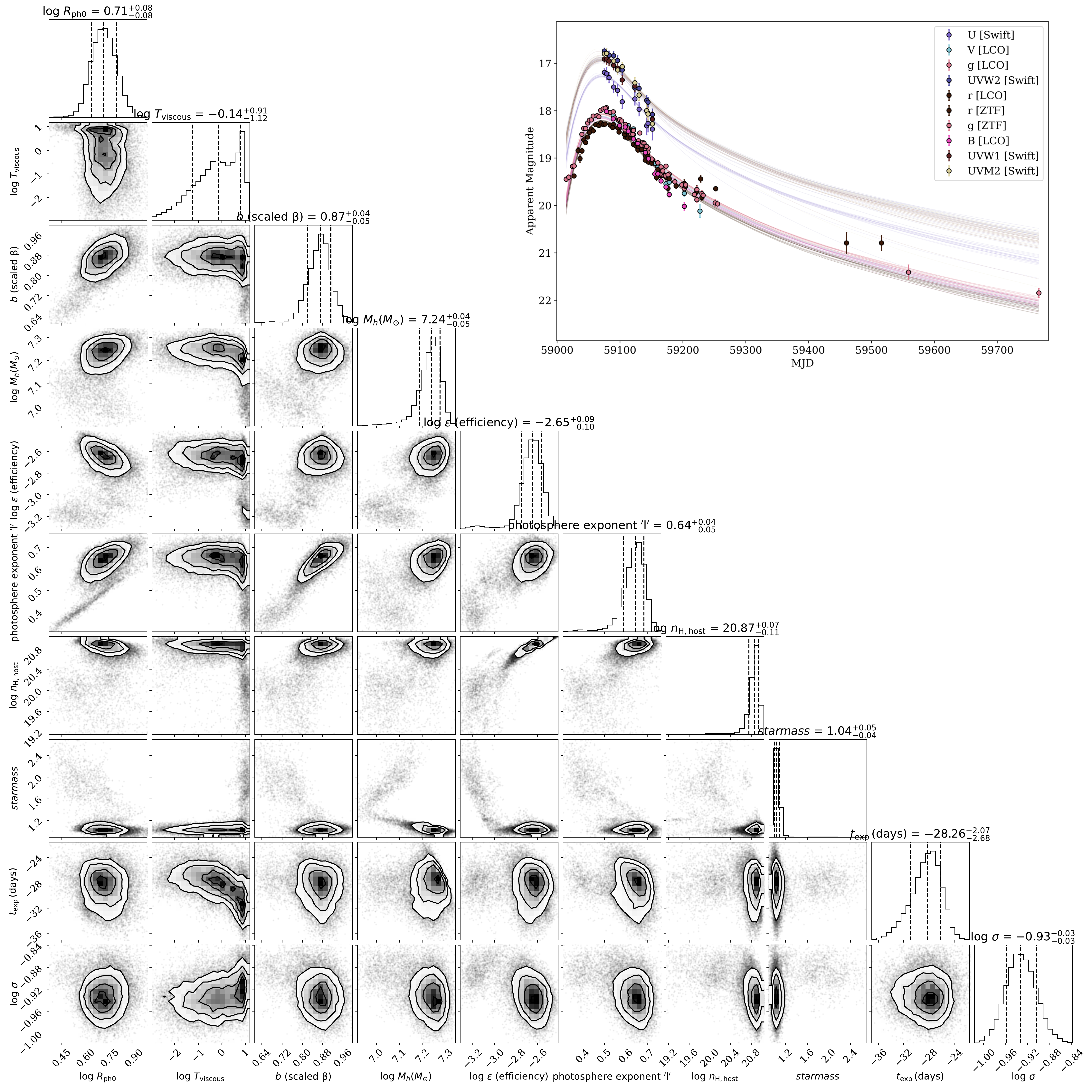}
    \caption{Corner plot and posterior light curves (upper right hand inset) from MOSFiT showing the convergence of parameter estimates based on fits to the UV/optical light curve of AT\,2020mot.}
    \label{fig:mosfit_lc}
\end{figure*}

One of the promises of tidal disruption observations is the use of the light curve to discern the mass of the central SMBH and the mass of the star before it was disrupted. Not only could such measurements serve as an alternative BH mass estimator to compare and improve existing BH mass relations, but larger samples of TDEs, especially from quiescent galaxies, can be used to generate population statistics on black hole masses in inactive galaxies contrasting with AGN, as well as inform the stellar mass distribution near galactic centers. However, as mentioned in Section \ref{sec:lightcurve}, the inferred values depend on the mechanics sourcing the flare, leading to different mass calculations based on assumptions of accretion or disk formation.

Two computational tools are publicly available to determine the mass of the SMBH and pre-disruption star. The Modular Open Source Fitter for Transients, MOSFiT\footnote{\url{https://github.com/guillochon/MOSFiT}} \citep{Guillochon2017}, has a TDE model that assumes fast circularization of the debris stream and subsequent formation of an accretion disk that is ultimately powering the event \citep[][]{Mockler2019}. Alternatively, TDEMass\footnote{\url{https://github.com/taehoryu/TDEmass}} uses only the peak bolometric luminosity and temperature to determine both SMBH and stellar mass by assuming the flare is produced from shocks on the debris' circularization path \citep[][]{Ryu2020}. Both methods were applied to our host-subtracted, extinction-corrected light curves which combine all Las Cumbres and Swift contributions.

We fit the MOSFiT model to the ZTF, Las Cumbres, and Swift combined data, which corresponds to the light curve rise, peak, and decline. We used the nested sampling mode and fixed redshift to the spectroscopically determined $z = 0.07$. MOSFiT pulls from hydrodynamical simulations of stellar tidal disruptions to model an event's multi-band luminosity evolution based on eight physical parameters beyond BH mass and stellar mass: photosphere power-law constant $R_{\text{ph}_{0}}$, viscous time delay $T_\text{viscous}$, impact parameter $\beta$, efficiency parameter $\epsilon$, photosphere power-law exponent $l$, column density of host $n_{H}$, and time of disruption relative to first observation $t_{\text{exp}}$ and a white noise parameter $\sigma$. The variation allowed in $\epsilon$ prevents MOSFiT from being firmly attached to the aforementioned prompt accretion model; low-efficiency parameters $\epsilon < 10\%$ could potentially correspond with expectations of stream-stream collisions contributing to rapid circularization and subsequent accretion \citep[][]{Jiang2016}. In Table \ref{table:fitting_masses} we show that MOSFiT finds a best-fit mass of the host galaxy's SMBH to be $M_{\text{BH}} = 17.37^{+1.66}_{-1.90} \times 10^6 M_{\odot}$ and the mass of the star before disruption $M_{*} = 1.04^{+0.05}_{-0.04} M_{\odot}$. These results are in agreement with \cite{Hammerstein2023} which finds $M_{\text{BH}} = 2.95-6.91 \times 10^6 M_{\odot}$, where the range is from the errors on their results from both MOSFiT and TDEMass. \cite{Yao2023} also find comparable results for AT 2020mot's host black hole mass, $M_{\text{BH}} = 2.09 - 10 \times 10^6 M_{\odot}$, using their measured velocity dispersion of the host galaxy and the \cite{KHo2013} relation.

\begin{figure}[t]
    \includegraphics[scale=0.5]{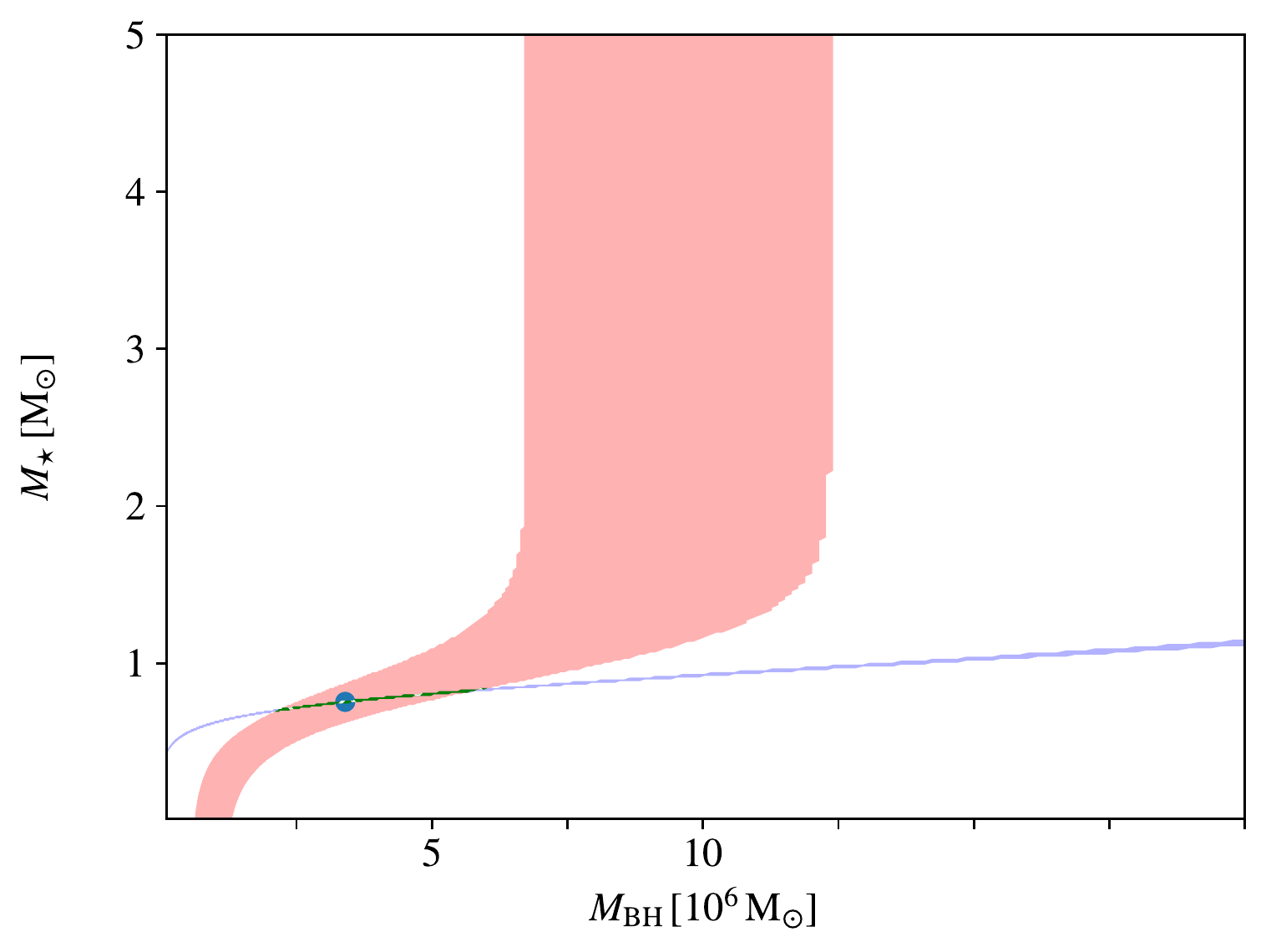}
    \caption{The output SMBH mass and star mass from TDEMass shown alongside their degeneracies. The blue line shows the solutions for masses from the given peak luminosity and the red strip shows the solutions from the peak temperature. The X-hatched green region thus indicates the overlapping solutions.}
    \label{fig:tdemass}
\end{figure}

TDEMass only uses the estimated bolometric luminosity and temperature at peak, instead of the entire observed light curve, to constrain SMBH and stellar mass estimates. By assuming slow circularization and interaction shocks of orbiting debris with incoming debris, the resulting energy dissipation equations directly give both BH and stellar masses from these two observables \citep[][]{Ryu2020}. In this scenario, AT\,2020mot's brightest measured bolometric luminosity ($3.05 \pm 0.2 \times 10^{43}$ erg/s) and corresponding temperature ($14600 \pm 2500$ K) result in an estimated $M_{\text{BH}} = 3.40^{+2.7}_{-1.4} M_{\odot}$ and $M_{*} = 0.75^{+0.09}_{-0.07} M_{\odot}$. The SMBH mass derived from these very different methods are just consistent with one another, though the stellar masses predicted by each have error estimates outside of the other's range.
\begin{deluxetable}{cccc}[t!]
\label{table:fitting_masses}
    \tablehead{
    \colhead{Fitting Tool} & \colhead{Epochs Used} & \colhead{$M_{\text{BH}}$ $(10^{6} M_{\odot})$} & \colhead{$M_{*}$  $(M_{\odot})$}
    }
    \startdata
         MOSFiT & Peak + Decline & $17.37^{+1.66}_{-1.90}$ & $1.04^{+0.05}_{-0.04}$\\
         TDEMass & Peak & $3.40^{+2.7}_{-1.4}$ & $0.75^{+0.09}_{-0.07}$
    \enddata
    \caption{Black hole masses estimated from MOSFiT and TDEMass.}
    \label{tab:bh_masses}
\end{deluxetable}

A delayed accretion scenario should coincide with X-ray brightening after the optical peak \citep[e.g.][]{Liu2022, Chen2022}. AT\,2020mot has only non-detections down to $1.33 \times 10^{43}$ erg/s even 18 months after peak (see Section \ref{sec:swift}). Furthermore, \cite{Piran2015} and \cite{Lu2020} find that self-intersecting stream collisions accelerate the circularization onto the accretion disk. In this case, the optical light curve observed is the product of the accretion flare being reprocessed through the outflow material in addition to emission from the collisions, all occurring at the same radii and temperature ($10^{-4}$ pc, $10^{4}$ K). Thus collisional shocks and reprocessed emission can both explain the blackbody parameters inferred for optical TDEs without X-ray detection. In either explanation, the X-rays from the inner accretion disk would remain undetected as they are behind the reprocessing layer of debris \citep{Dai2018}, which may explain our lack of detection, although it is also possible that accretion may not have started yet or our detection limits were not deep enough to observe the accretion flare.

This competition between radiatively efficient reprocessing versus inefficient shocks is dependent on the density of the gas distribution along the tidal streams, the angle between the orbiting and impacting streams upon collision, and whether collisionally induced outflows are launched in spherical geometries \citep[][]{Bonnerot2021}. \cite{Liodakis2022} find a peak intrinsic polarization degree of $25\pm4$\% for AT 2020mot which they claim can only be explained by multiple stream shocks each at differing polarization angles. TDEmass and MOSFiT disagree by an order of magnitude on the SMBH mass and the mass of the star pre-disruption, so the effects of each emission mechanism produces notably different properties from the same light curve. We investigate the strength of these black hole mass estimates by comparing against host galaxies properties in the following section.

\section{Host Galaxy Properties} \label{sec:galaxy}

The Las Cumbres host spectrum, shown in Figures \ref{fig:spectra} and \ref{fig:spectra_warped}, shows Na I D $\lambda\lambda5890, 5896$, Mg Ib $\lambda5175$, and Ca H $\&$ K $\lambda\lambda3934, 3968$ absorption lines. The relationship between host extinction and the equivalent width of the Na I D doublet has been well established \citep[][]{1997A&A...318..269M}, but our spectrum is too noisy and lacks the resolution to discern the equivalent width \citep{2012MNRAS.426.1465P}, so we do not estimate host reddening from this process. The spectrum also confirms a galactic environment with a young stellar component from the Balmer absorption lines characteristic of A stars, which gives rise to the Balmer lines, yet no emission, particularly no [O III] expected from active star formation. The presence of young stars without ongoing star formation is indicative of a recently-quenched galaxy, such a post-starburst or post-merger E+A galaxy \citep[][]{Zabludoff1996}, as found to be the case in many TDEs \citep{Arcavi2014, 2016ApJ...818L..21F}. Furthermore, the detection of Mg Ib $\lambda5175$ and Ca H $\&$ K $\lambda\lambda3934, 3968$ absorption lines in the host spectrum reflects an older stellar population that is simultaneously present. \cite{Liodakis2022} also determine the galaxy to be spheroidal and likely elliptical. \cite{KK04} consider elliptical galaxies to be dominated by a classical bulge, thus we expect the stellar mass of the host to be comparable to the bulge mass.

We seek to assess how SMBH mass relations via host galaxy properties compare with analyses from TDE light curves (see Section \ref{sec:lightcurve}). We pursue this analysis by fitting the available photometric data of the host galaxy WISEA J003113.52+850031.8 \citep[][]{Skrutskie2006}, as listed in Table \ref{tab:hostphot}, with the SED fitting code \texttt{BAGPIPES} \citep[][]{Carnall2018}. For further detail on our use of this fitting tool, see the Appendix. 
\begin{deluxetable}{cccc}[t!]
\label{table:gal_phot}
    \tablehead{
    \colhead{Filter} & \colhead{Source}& \colhead{Magnitude} & \colhead{Flux} \\
    \colhead{} & \colhead{} & \colhead{} & \colhead{$\mu$Jy}
    }
    \startdata
         UVW2 & Swift & 20.19 & 6.88\\
         UVM2 & Swift & 19.83 & 9.57\\
         UVW1 & Swift & 19.55 & 15.66\\
         B & Las Cumbres & 18.84 & 119.705 \\
         g & PAN-STARRS & 18.46 & 154.51\\
         V & Las Cumbres & 17.90 & 243.215\\
         r & PAN-STARRS & 17.73 & 300.47\\
         i & PAN-STARRS & 17.34 & 435.60\\
         J & 2MASS & 15.54 & 966.0\\
         H & 2MASS & 14.37 & 1180.0\\
         K & 2MASS & 14.47 & 1090.0\\
         W1 & WISE & 13.93 & 826.0\\
         W2 & WISE & 13.89 & 429.0
    \enddata
    \caption{Photometry of the host galaxy WISEA J003113.52+850031.8 used in fitting with \texttt{BAGPIPES}. All magnitudes are given in the Vega system except for $gri$ magnitudes, given in the AB system.}
    \label{tab:hostphot}
\end{deluxetable}

The mass formed from the best-fit double power law is $\text{log}_{10}(M_{*} / M_{\odot}) = 10.02^{+0.05}_{-0.03}$. If taking the stellar mass formed to equate the bulge mass, we can use relations between the bulge mass and SMBH mass to cross-check the estimates of the SMBH mass from Section \ref{subsec:mosfit}.  We start with the $M_{\text{BH}} - M_{\text{bulge}}$ relation from \cite{Haring2004} which gives
\begin{equation}
\label{eq:Haring04}
    \text{log}_{10} \left (\frac{M_{BH}}{M_{\odot}} \right ) = 8.20  +  1.12\text{log}_{10} \left ( \frac{M_{bulge}}{ 10^{11} M_{\odot}} \right )
\end{equation}
derived using 30 nearby galaxies, finding most SMBHs to fall between $10^{7}$ - $10^{9} M_{\odot}$. \cite{Scott2013} evolved this relation for lower-mass galaxies (and subsequently lower-mass central BHs), finding
\begin{equation} 
\label{eq:Scott13}
    \text{log}_{10} \left (\frac{M_{BH}}{M_{\odot}} \right ) = 2.22  +  7.89\text{log}_{10} \left ( \frac{M_{bulge}}{3 \times 10^{10} M_{\odot}} \right ).
\end{equation}
Finally, we also consider the relation developed specifically for TDE host galaxies in \cite{Ramsden2022},
\begin{equation} 
\label{eq:Ramsden22}
    \text{log}_{10} \left (\frac{M_{BH}}{M_{\odot}} \right ) = 7.0 +  0.18\text{log}_{10} \left ( \frac{M_{bulge}}{10^{11} M_{\odot}} \right )
\end{equation}
which was derived using a sample of 32 TDEs whose BH masses were determined with MOSFiT only \citep[][]{Nicholl2022}.

We consider all three above relations and compare their estimates with each other, as well as with our results from MOSFiT and TDEmass in Section \ref{subsec:mosfit}. For the \cite{Haring2004} relation the BH mass is found to be $M_{BH} = 12.65^{
+1.25}_{-1.25} \times 10^6 M_{\odot}$; for \cite{Scott2013} $M_{BH} = 7.50^{+1.63}_{-1.34} \times 10^6 M_{\odot}$; for \cite{Ramsden2022} $M_{BH} = 6.66 \times 10^6 M_{\odot}$.

The value determined from the TDE-specific Equation \ref{eq:Ramsden22} falls between that found from our own MOSFiT run and TDEMass fit to the light curve of AT2020mot. However, the relation was itself calibrated to MOSFiT-determined BH masses. Equation \ref{eq:Scott13} also matches the MOSFiT-derived BH mass within the errors. Given most BH mass relations still steer toward higher-mass black holes found in galaxy surveys than those probed by TDE studies, it is also unsurprising that BH mass estimated from Equations \ref{eq:Haring04} and \ref{eq:Scott13} may ultimately overestimate the masses of TDE black holes. In the same vein, the bulge mass may be overestimated from tools like \texttt{BAGPIPES} whose models are based off of galaxy samples of higher masses, and by assuming the stellar mass is equal to the bulge mass.
\begin{deluxetable}{ccc}[t!]
\label{tab:bhmass_galaxyrelations}
    \tablehead{
    \colhead{Relation} & \colhead{Source}& \colhead{$M_{BH}$} \\
    \colhead{} & \colhead{} & \colhead{$10^{6} M_{\odot}$}
    }
    \startdata
         $M_{\text{BH}} - M_{\text{bulge}}$ & \cite{Haring2004} & $12.65 \pm 1.25$\\
         $M_{\text{BH}} - M_{\text{bulge}}$ & \cite{Scott2013} & $7.50^{+9.13}_{-6.16}$\\
         $M_{\text{BH}} - M_{\text{bulge}}$ & \cite{Ramsden2022} & 6.66\\
         $M_{\text{BH}} - L_{V}$ & \cite{DeGraf2015} & $3.17 \pm 0.04$\\
         $M_{\text{BH}} - L_{V}$ & \cite{KHo2013} & $1.36 \pm 0.38$\\
    \enddata
    \caption{Estimates of BH mass from relations using host galaxy properties.}
\end{deluxetable}

Another estimate for SMBH mass uses a scaling with the host galaxy's luminosity. Using Las Cumbres template images as the source for $V$-band aperture photometry of the host (listed in Table \ref{tab:hostphot}), we also estimate the BH mass via the mass-luminosity relation
\begin{equation}
\label{eq:DeGraf2015}
    \text{log}_{10} \left (\frac{M_{BH}}{M_{\odot}} \right ) = 8.249 + 1.062\text{log}_{10} \left ( \frac{L_{V,bulge}}{10^{10.5} L_{\odot}} \right )
\end{equation}
from \cite{DeGraf2015}, using the best-fit values to their simulations at $z = 0.06$. The host $V$-band luminosity, $2.73 \times 10^{42}$ erg/s, gives the estimated $M_{BH} = 3.17^{3.21}_{3.13} \times 10^{6} M_{\odot}$. This estimate is the lowest of all other calculations using host galaxy properties. It is worth noting that Equation \ref{eq:DeGraf2015} is developed from a sample of simulations and not observations; this allows for a large sample but is not verified by observational data. To make use of a relation motivated by empirical data, we turn to Equation 6 from \cite{KHo2013},
\begin{equation}
\label{eq:KHo2013}
    \text{log}_{10} \left (\frac{M_{BH}}{M_{\odot}} \right ) = 8.734 + 1.21\text{log}_{10} \left ( \frac{L_{K,bulge}}{10^{11} L_{K,\odot}} \right )
\end{equation}
in which we use archival Two Micron All Sky Survey \citep[2MASS][]{2006AJ....131.1163S} images of the host galaxy to determine the host $K$-band luminosity $2.16 \times 10^{42}$ erg/s, which gives $M_{BH} = 1.36 \times 10^{6} M_{\odot}$, again a lower estimated value compared to the $M_{BH} - M_{bulge}$ relations.

Among the relations to estimate a SMBH mass via observed host galaxy properties, the relations dependent on bulge mass estimate a heavier SMBH in WISEA J003113.52+850031.8 than the relations dependent on luminosity. The former are closer in scale to the mass determined from AT2020mot's photometry via MOSFiT, while the latter are notably close to the estimate from TDEmass.

\section{Circumnuclear Dust Models} \label{sec:dust}

The brightness of AT2020mot in the $i$-band exceeds that of other optical TDEs (see Figure \ref{fig:colors}) and shows a re-brightening bump after MJD 59160, starting 83 days after maximum light. We explore whether models of circumnuclear dust can explain the observations as a ``dust echo" of delayed, reprocessed emission. Past works on potential dust echoes in TDEs have relied almost solely on WISE data, and none discuss the possibility of an echo being observed in the $i$-band \citep[e.g.][]{JiangN2016, Dou2016, VanVelzen2016, Jiang2017, Li2020, Stein2021, Jiang2021, Wang2022, Onori2022}.

We use the methods of \cite{VanVelzen2016} to model the response light curve in infrared bands from dust grain absorption of UV to optical light. See Sections 2 and 4 of \cite{VanVelzen2016} for specifics of the response function based on assumptions of the type and size of dust grain, and the parameterization of dust geometries around the SMBH. Assuming spherically symmetric TDE emission and a graphite\footnote{Silicate grains have a much lower sublimation temperature such that any present would not last to reprocess the emission from a UV/optical flare.} dust grain of radius $a \sim 0.1 \mu m$, we begin with the generic model,
\begin{equation}
\label{eq:VV16_generic}
    L_{\nu, \text{echo}} \propto \int \Psi(\tau) L_{\text{TDE}}(t-\tau) d\tau
\end{equation}
where $\Psi$ is the response function of the dust and $L_{\text{TDE}}$ is the integrated UV and optical luminosity measured at time $t - \tau$, with $t$ being the time of IR detection, and $\tau$ being the delay between when the TDE light was emitted and when it was reprocessed to IR based on the distance $R$ and angle $\theta$ of the dust from the TDE. \cite{VanVelzen2016} parameterize $\tau$ with respect to a polar axis aligned to the observer's line of sight, such that
\begin{equation}\label{eq:tau}
    \tau = \frac{R}{c}(1 - \cos\theta).
\end{equation}
With this framework, the light curve expected from a thin spherical shell of dust is modeled as
\begin{equation}
\label{eq:VV16_sphere}
    L_{\nu, \text{echo}} = A C_{q} B_{\nu}^{'}2\pi \int_{0}^{\pi} L_{\text{TDE}}(t-\tau)\sin \theta d\theta
\end{equation}
where $A$ is a constant of amplification reflecting the ratio between observed IR luminosity and the amount calculated from reprocessing, and $B_{\nu}^{'}$ is a modification of the Planck function $B_{\nu}$ which encodes the IR-specific response to the absorbed UV and optical luminosity, such that $B_{\nu}^{'} = B_{\nu}\nu^{q}$ \citep{VanVelzen2016}. Maintaining the assumption of dust grains of size $a \sim 0.1 \mu m$, \cite{Draine1984} gives $q=1.8$ and $C_{q}$ is a constant of normalization $C_{q} = 1/\int B_{\nu}^{'} d\nu$.

This model can be adjusted for including shell thickness by including integration over extended radii:
\begin{equation}
\label{eq:VV16_thicksphere}
\begin{split}
    L_{\nu, \text{echo}} = & C_{q} 2\pi \int_{R_0}\int_{0}^{\pi} L_{\text{TDE}}(t-\tau)\sin \theta \\ & \times  A \times n(R) B_{\nu}^{'}(T(R)) d\theta dR.
\end{split}
\end{equation}
The amplification factor $A$ remains in the integral in cases of radial thickness because it can vary at each radius. We can thus also evaluate the thick shell models with varying densities $n(R)$ of dusty material at each radius step, and account for temperature variation with increasing radius as $T(R) = T_{0} (R/R_{0}) ^{-0.345}$ \citep[][]{VanVelzen2016}.

\begin{figure}[t]
    \centering
    \includegraphics[scale=0.5]{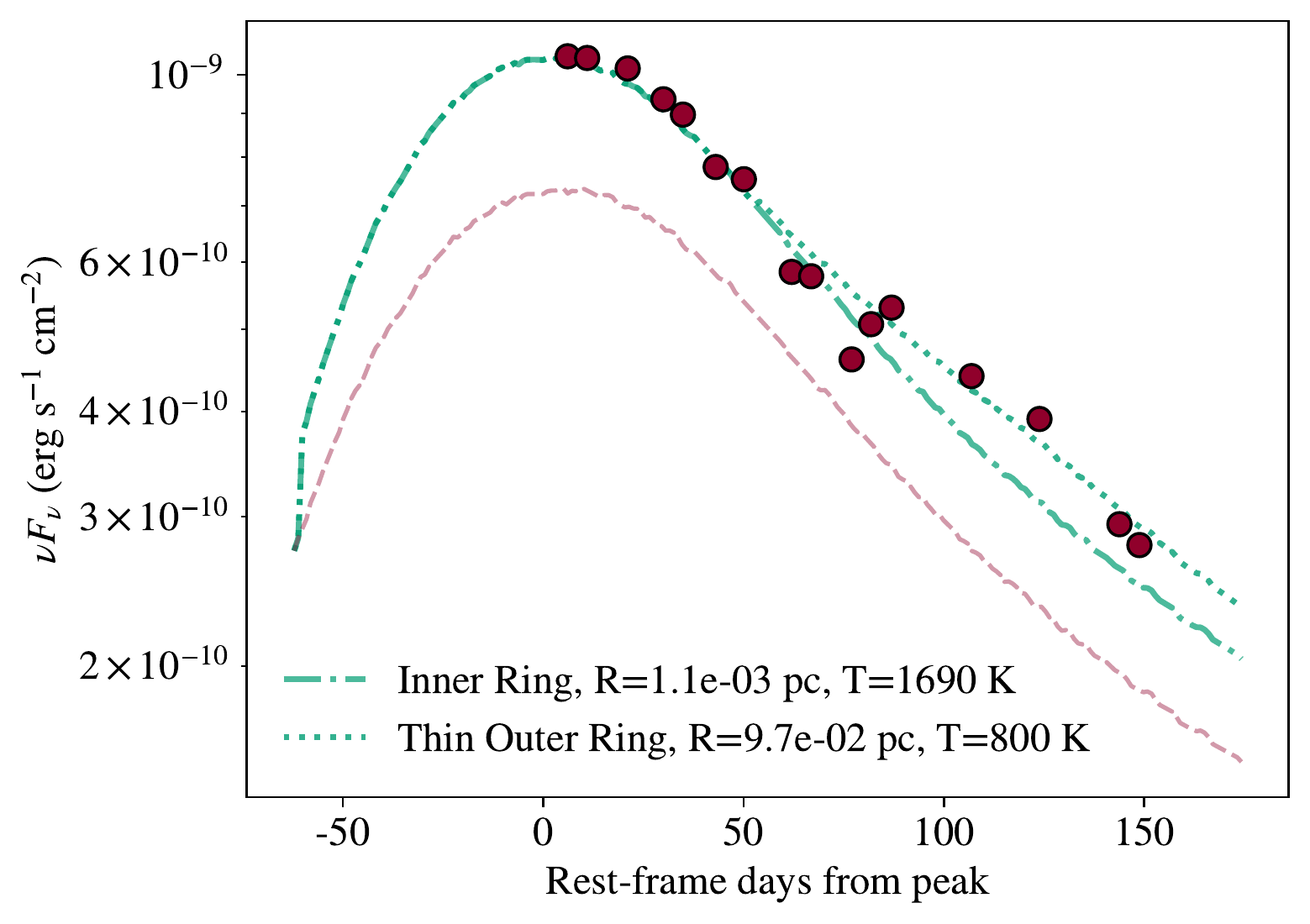}
    \caption{The observed $i$-band light curve (red-filled circles) compared with the estimated intrinsic emission (red dashed lines) alongside the inclusion of dust re-emission (blue dashed and dash-dot lines). The best-fit light curves from dust reprocessing are a thin inner ring at 0.001 parsecs (dash-dotted line) from the central supermassive black hole, and another ring at 0.09pc (dashed line).}
    \label{fig:bestfit_dust}
\end{figure}

Furthermore, by restricting the angular variation, we can mimic a ring (all dust at one inclination angle and radius), thick ring (superposition of rings at increasing radii), and torus geometry (small central segment of the spherical shell at one inclination angle and superimposed radii). For instance, a face-on thick ring would be modeled with
\begin{equation}
\label{eq:VV16_thickdisk}
\begin{split}
    L_{\nu, \text{echo}} = &  2\pi \int_{R_0} A C_{q} n(R) B_{\nu}^{'}(T(R)) L_{\text{TDE}}(t-\tau) dR.
\end{split}
\end{equation}
because all IR emission follows the same time delay without angular dependence such that $\tau = R/c$ in the face-on case. By extension, a ring or torus at a different inclination would also integrate over angular dependencies.

We test six fiducial geometries of dust: thin spherical shell, thick spherical shell, face-on thin ring, face-on thick ring, face-on thin torus, face-on thick torus. The cases of radial thickness are also tested with both constant and Bondi density ($n(R) \propto R^{-3/2}$) profiles. We then perform MCMC fitting to estimate the best-fit combination of distance $R$ from the SMBH (with log priors over $1e14$ -- $1e18$ cm), its temperature of re-emission $T$ (flat priors over 100--5000 K), and the amplitude of reprocessing (flat priors over 0.001--10) for each configuration of dust whose re-emission could match and the observations. The $i$-band data is separated into ``pre-bump" (MJD $< 59150.0$) and ``bump" (MJD $>= 59150.0$) groups. We begin with pre-bump data, and make the significant assumption that the intrinsic $i$-band luminosity follows what would be expected with the application of the PS1-10jh template \citep{Gezari2012}, shifted to match the luminosity difference found for AT2020mot between $g$- and $r$-bands. This assumed intrinsic luminosity can be seen in Figure \ref{fig:bestfit_dust} and is subtracted from the observed pre-bump luminosity measurements, to isolate the light that is in excess from the TDE flare and thus likely a result of dust reprocessing.

\begin{figure*}[t]
    \centering
    \includegraphics[scale=0.6]{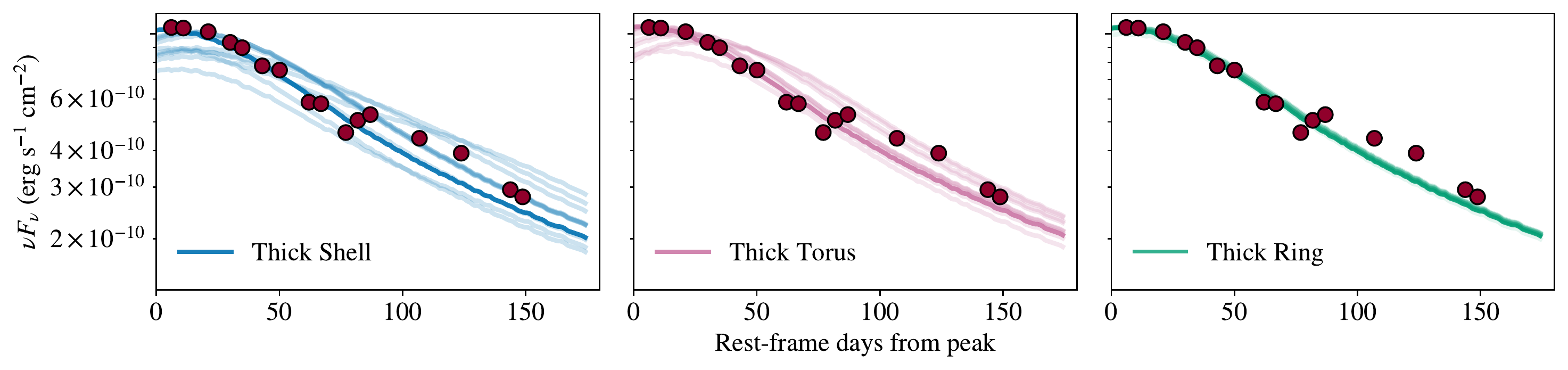}
    \caption{Posteriors of the reprocessed emission from different thick-dust geometries with Bondi density profiles alongside the $i$-band excess (shown in red-filled circles). The left panel shows the posteriors for a shell in blue lines; middle panel shows those of a torus in red lines; the right panel shows those of a ring in green lines. Varying radius directly delays the onset of the enhanced emission, while varying temperature affects the amplitude of the reprocessing.}
    \label{fig:prebump_geometries}
\end{figure*}
\begin{figure*}[t]
    \centering
    \includegraphics[scale=0.6]{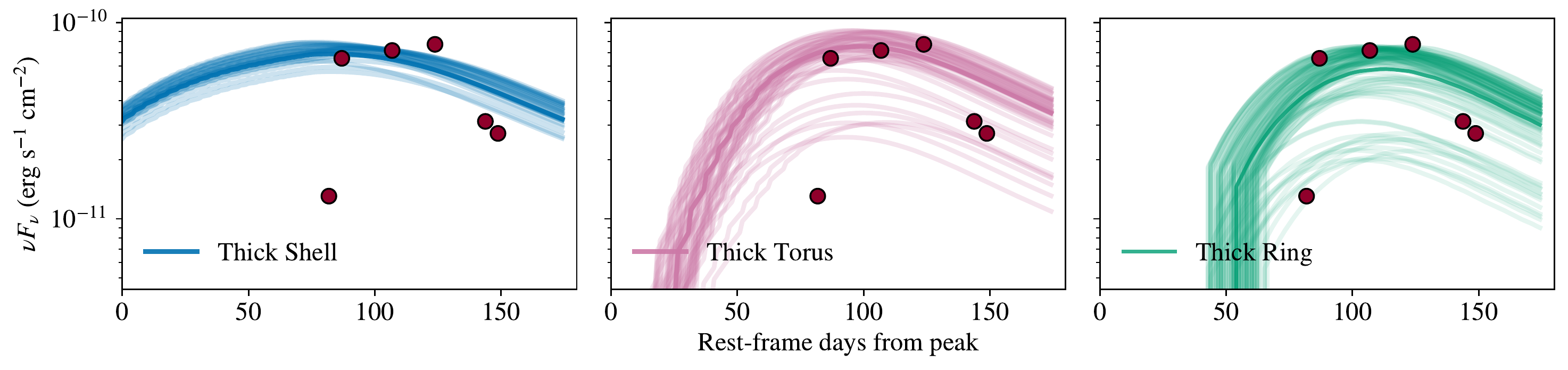}
    \caption{Same as Figure \ref{fig:prebump_geometries}, but for the photometry points and dust model fitting only correspond to the bump in $i$-band brightness beginning after 80 days from the optical peak.}
    \label{fig:bump_geometries}
\end{figure*}

\begin{figure}[t]
    \centering
    \includegraphics[scale=0.4]{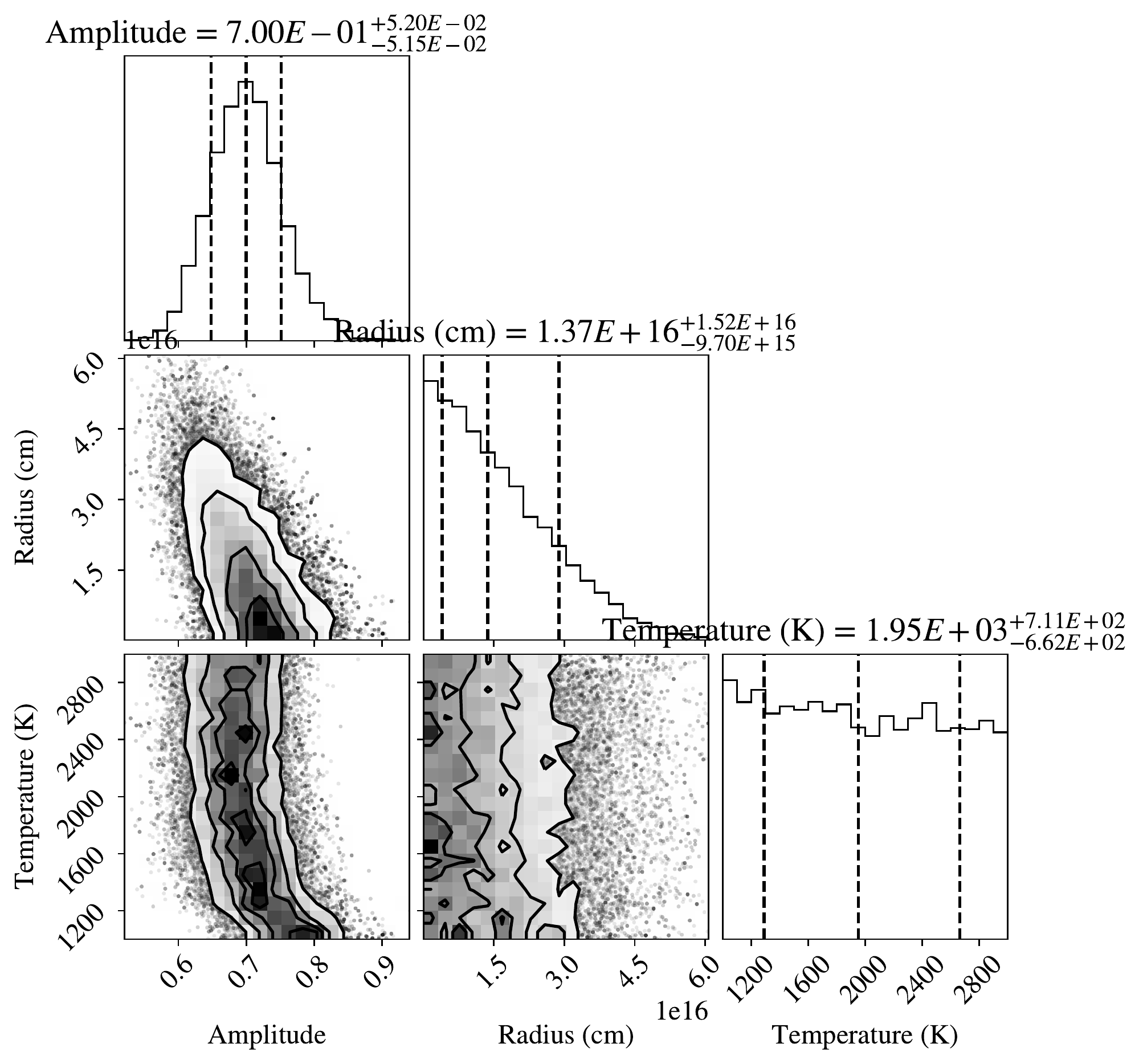}
    \caption{Corner plot showing the convergence of the ring dust model's fitting parameters for the excess in $i$-band brightness at early times.}
    \label{fig:corner_early}
\end{figure}
\begin{figure}[t]
    \centering
    \includegraphics[scale=0.4]{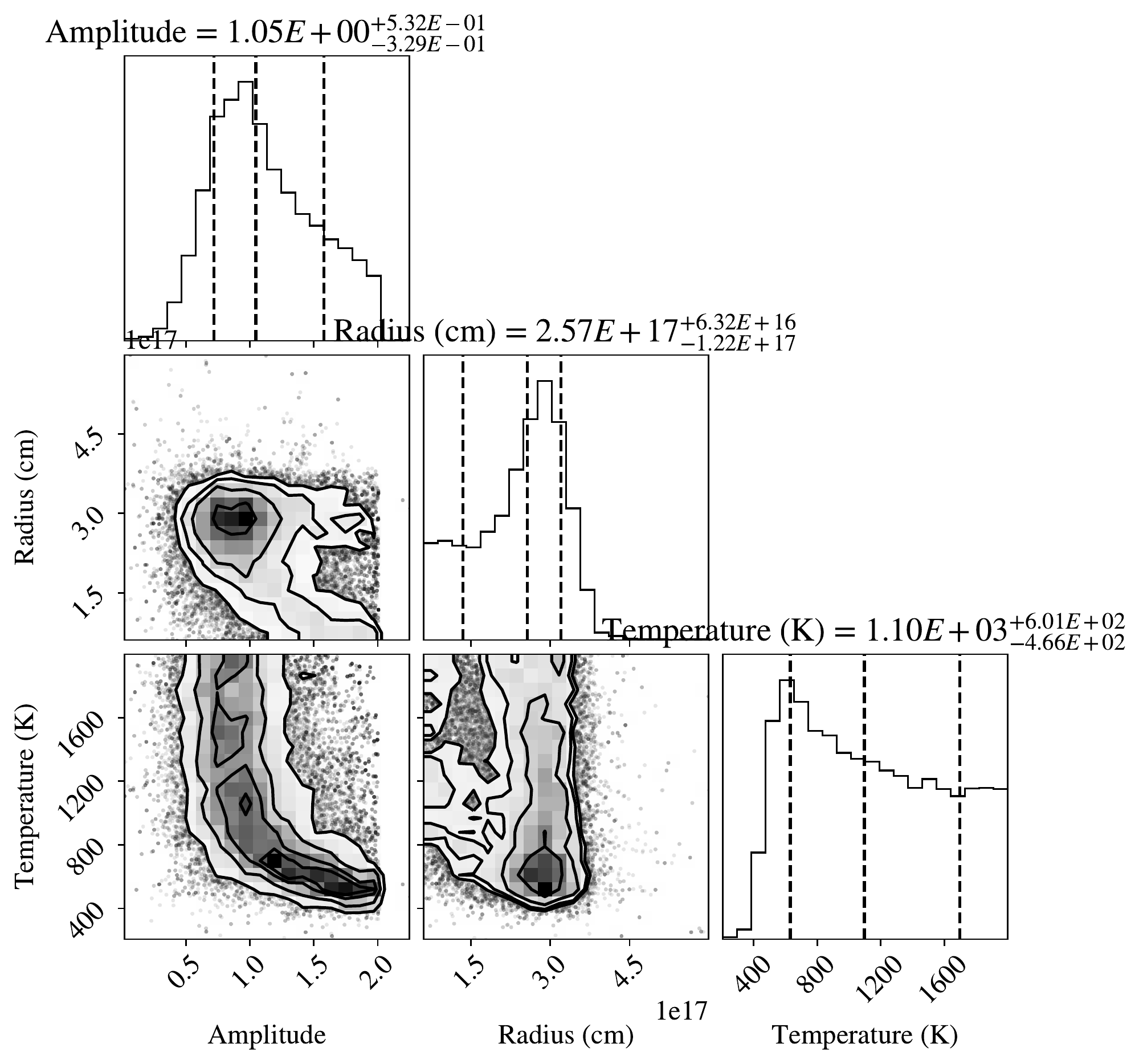}
    \caption{Corner plot showing the convergence of ring dust model's fitting parameters for the bump in $i$-band brightness beginning after 80 days from the optical peak.}
    \label{fig:corner_bump}
\end{figure}

The $i$-band light curve is consistent with the expected intrinsic TDE emission combined with reprocessed emission of two face-on rings each with a Bondi density profile. The innermost radius of the first ring is $R_\text{sub} = 1-5\times 10^{-3}$ pc with a temperature of $T \approx 1700 K$, and the radius of the second ring is at $R_\text{sub} = 9.7\times 10^{-2}$ pc with a temperature of $T \approx 800$ K, as shown in Figure \ref{fig:bestfit_dust}. These values are determined from the results with the highest likelihood scores after MCMC fitting, instead of using the median of the fit, because of the varied degree of convergence for each parameter. For instance, Figure \ref{fig:corner_early} shows that the inner ring of dust is matched by the smallest of scales such that the radius solution does not converge at the medium value, while the temperature solution is not converged and results are unaffected by temperature variance.

In Figures \ref{fig:prebump_geometries} and \ref{fig:bump_geometries}, the results from this fitting process for different geometry examples are presented for both the pre-bump and bump datapoints. There is little difference between the scenarios of a spherical shell, ring, or torus at similar scales in generating enough reprocessed light to match the observations before the bump. The bump, however, is not uniformly fit by all geometries. The spherical shell models produce flatter, wider reprocessing light curves (described as ``square waves" in \citealt{VanVelzen2016}) due to the different distances at which light will have to travel from the origin at the TDE to the surrounding dust at different angles. The more centrally located the dust is along the plane perpendicular to our line of sight (that is, the closer to a face-on ring or torus), the more the reprocessing emission is received as a steeper and sharper light curve. As this qualitatively describes a bump, we expected ring and torus configurations to be responsible for the bump, and confirm as follows.

\begin{figure}[t]
    \centering
    \includegraphics[scale=0.45]{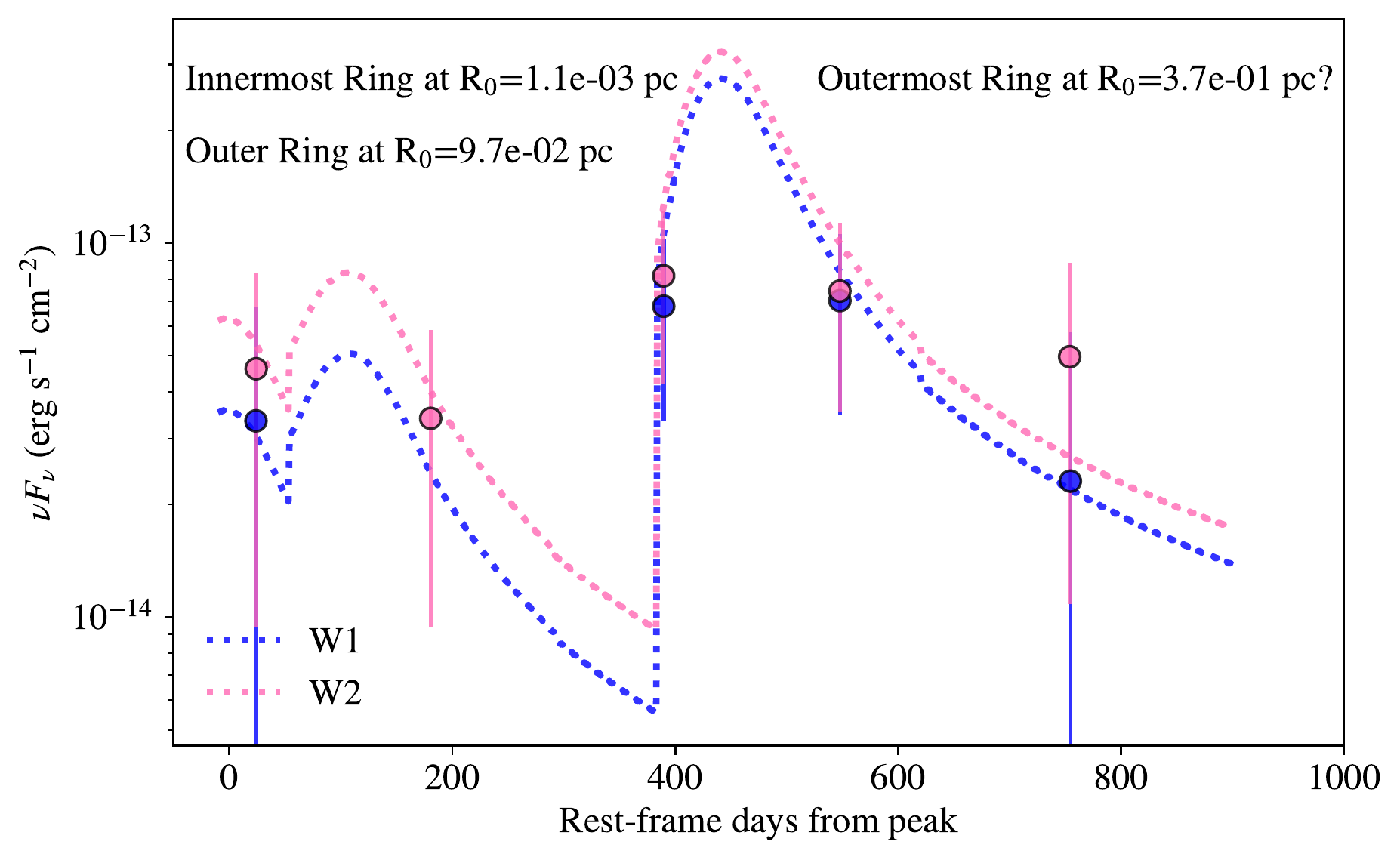}
    \caption{The reprocessing signatures of the inner ($R = 0.001$pc and outer $R = 0.09$pc dust rings, which best-fit the $i$-band data, in $W1$ (blue) and $W2$ (pink) alongside the subtracted WISE light curves for AT 2020mot. The first two rings produce signatures which overlap within the errors of the $W2$ data, and overlap with the first epoch of $W1$ data. We also show that the latter 3 epochs of both WISE bands can be fit with a third, outermost ring at $R=0.37$pc.}
    \label{fig:WISE_3rings}
\end{figure}

To fit the luminosity of the bump that is only from a secondary dust reprocessing, the luminosity from the first dust grouping (sphere, ring, torus) is subtracted from the observed bump points. We test only superpositions of like geometries: inner and outer shell; inner and outer ring; inner and outer torus. Figure \ref{fig:bestfit_dust} also shows the best-fit light curve to the bump, which is provided by an outer ring at $R=0.09$ pc with a lower temperature than the inner ring. Inner and outer torus configurations were best-fit at the same radii, however, the outer torus produces an ultimately flatter light curve than the outer ring, leaving the ring to match the observed bump best, as shown in Figure \ref{fig:bump_geometries}. No spherical shell model was found to fit the bump data, as the flatter reprocessing light curve from a sphere would also contribute luminosity to the pre-bump phase. Concentric spherical shells are thus unlikely to explain the early $i$-band excess and late-time bump.

We show the effect of our best-fit concentric rings of thin dust in the mid-infrared by showing the reprocessing signatures in the WISE $W1$ and $W2$ bands alongside the host-subtracted WISE epochs of AT 2020mot in Figure \ref{fig:WISE_3rings}. Only two epochs overlap in time with the expected reprocessing output of the two rings, and the signatures are indeed within the errors of both of these epochs. Furthermore, we show that a third ring at $R = 0.37$pc can also reproduce the last three epochs of WISE photometry within errors.

By calculating the dust covering factor $f_{c}$ as $L_{\text{dust, max}}/L_{\text{TDE, max}}$ where $L_{\text{TDE}}$ is the maximum luminosity of the TDE integrated over UV and optical wavelengths, we also find an inner ring covering factor of $f_{c} = 4.5\%$ and an outer ring covering factor of $f_{c} = 0.82\%$.

We explore the physical sensibility of these models and the implications of low covering factors in Section \ref{sec:Discussion}.

\section{Discussion}\label{sec:Discussion}

We discuss the implications of two possible explanations for the $i$-band excess and bump in AT 2020mot: dust echoes and extended red emission (ERE).

\subsection{Dust echoes: Concentric Rings?}

The dust fits to the $i$-band, in addition to the best-fit model's consistency with WISE data, make AT 2020mot the first TDE with multi-wavelength signatures consistent with concentric rings of dust, as well as the first TDE with dust as close as $0.001$pc to a SMBH.

The $i$-band excess and bump has not yet been reported in other TDEs. Enhanced or late-time infrared emission in TDEs has been limited mostly to observations from WISE, whose sparse sampling compared to the lifespan of the TDE limits our ability to constrain the geometry of the IR-enhancing material. The Las Cumbres $i$-band is used simultaneously with optical follow-up to ensure thorough multi-band coverage of the light curve. This early and rapid NIR follow-up can uncover dust at previously-unreachable sub-parsec scales near a SMBH, as is the case with AT 2020mot.

\begin{figure}[t]\label{fig:schematic}
    \centering
    \includegraphics[scale=0.07]{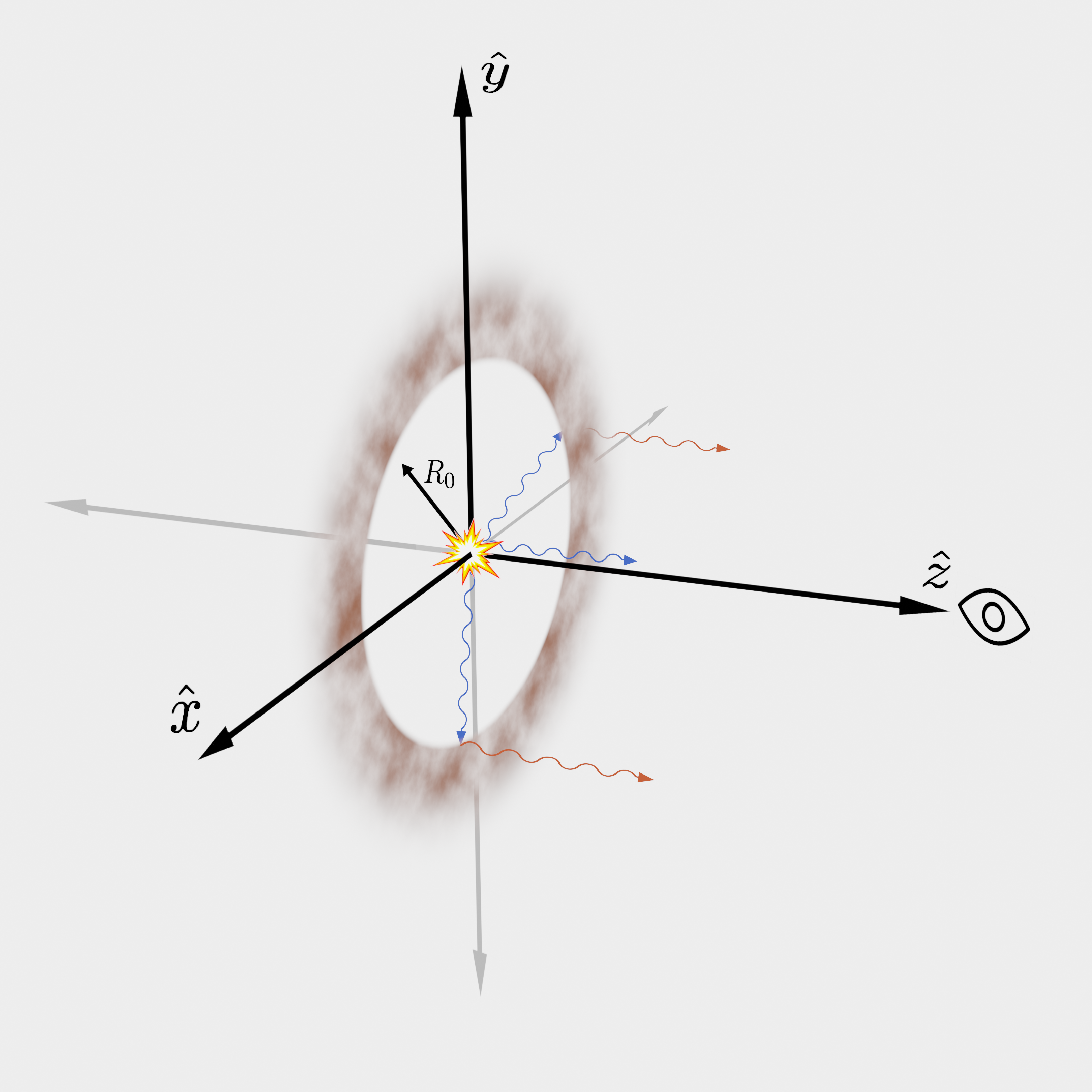}
    \caption{Schematic representation of the ring model of dust surrounding a SMBH, in which the dust lies mostly in the plane facing the observer. Light from the central TDE is emitted isotropically, thus the observer sees unobscured light from the flare along the line of sight. TDE light that is emitted perpendicular to the line of sight meets the ring of dust that is densest at the innermost radius $R_{0}$, and the density of dust radially outward decreases following a Bondi profile. This dust absorbs the UV and optical light, and dust that is not sublimated will re-process the absorbed light into the infrared, thus sending a second pulse of light to the observer in the IR only.}
\end{figure}

Our analysis of the $i$-band light curve finds that the pre-bump and bump phases are both well fit by two circumnuclear rings of dust. Face-on disks of dust are physically sensible phenomena to occur around supermassive black holes, although a torus is more commonly invoked, particularly for AGN \citep[][]{Barvainis1987, VanVelzen2016, JiangN2016}. When discussing common dust configurations around SMBHs, it is important to note differences in scale and mass of the dust as typically found via reverberation mapping. AGN tori are usually found on scales around $\sim0.01-1$ pc from the central SMBH, but with a wide range in outer radius extent \citep[][]{2014ApJ...788..159K, 2019ApJ...886..150M}. Disks and rings, however, are more likely in cases with low dust covering factors, because a torus needs a high accretion rate to be sustained \citep[][]{VanVelzen2016}. Such disks, if face-on, would reprocess oncoming UV and optical light with equal delay times at every azimuthal angle, producing a pulse of enhanced near-infrared light. If close enough to the initial flare near the SMBH, this light would add to the intrinsic light of a TDE's decline in a corresponding IR filter.

We find concentric rings at $R_{1} = 0.001$ pc and $R_{2} = 0.09$ pc (see Figure \ref{fig:bestfit_dust}) simultaneously fit the pre-bump and bump light curves. Our models give a maximum dust luminosity of $5.38 \times 10^{41}$ erg/s for the inner ring, and $9.75 \times 10^{40}$ erg/s for the outer ring. Furthermore the covering factors of $4.5\%$ and $0.82\%$ at each ring are in agreement with other TDE dust covering factors around $f_{c}\sim1\%$ \citep{Jiang2021}, and they are orders of magnitude lower than the covering factors found for typical AGN \citep{Barvainis1987}. \cite{VanVelzen2016} and \cite{JiangN2016} note that these low covering factors imply far too little dust for torus formation, and more likely correspond to geometrically thin configurations such as rings.

\subsection{Alternative Explanation: Extended Red Emission?}

Extended Red Emission (ERE) is a broad emission feature seen in some diffuse systems that is due to the absorption and re-emission of UV photons by unidentified dust grains, with measured observations peaking anywhere from $6000\angstrom$\ for the diffuse interstellar medium \citep{Szomoru1998} to $7500\angstrom$ for HII regions \citep{Sivan1993}. \citet{Smith2002} also found that the ERE peak wavelength increases with increasing densities of the radiation field that excites the dust. This correlation requires an evolving size distribution of the re-emitting dust particles by gradual photo-fragmentation reducing the number of smaller particles.

Extended red emission was reported for Sgr A* by \citet{Ghez2005}, posited as evidence of a dust cloud that would have otherwise been pulled into a ring around the SMBH by tidal interaction if it were too physically close; thus it was inferred that the dust was more likely a cloud along our line of sight to Sgr A*. ERE has also been found near other galactic nuclei such as NGC 4826 in HII regions matching the expectation of dust lanes \citep{2002ApJ...569..184P}.

There is little study on ERE in other galaxies, and much debate on the particles that produce the emission, with polycyclic aromatic hydrocarbons and nanodiamonds being among the top contenders \citep{Chang2006, Rhee2007}. However, the broad properties of extended red emission that strongly depend on environmental factors, such as the density of the dusty material and intensity of the radiation field, all make ERE a possible explanation for the anomalous $i$-band emission of AT 2020mot. Since the Las Cumbres $i$-band peaks near $7500 \angstrom$ the excess infrared emission of AT 2020mot is indeed similar to that observed for Sgr A* and NGC 4826.

Further studies are needed to assess whether the UV output of TDEs are adequate to explain ERE, and how close the candidate particles responsible for ERE can survive in galactic nuclei.

\section{Conclusions}\label{sec:conclusion}

AT2020mot is a UV/optical tidal disruption event in a galaxy at $z=0.07$. Its light curve is comparable to well-sampled optical TDEs such as PS1-10jh \citep{Gezari2012}, except in the $i$-band, which is more luminous than expected and shows an extra ``bump" in brightness along the decline. The host properties fit an ``E+A" or ``K+A" classification that is possibly post-starburst, and/or a product of past mergers \citep[][]{Zabludoff1996}. The association between E+A galaxies and TDEs has been well established since first reported in \cite{Arcavi2014}. The host galaxy properties, UV/optical light curve, lack of X-rays (see Figure \ref{fig:xray_lc}), and radio upper-limits of $27 \mu$Jy at 15 GHz as reported in \cite{Liodakis2022}, all make AT2020mot an otherwise typical optical TDE, if not for its $i$-band light curve peculiarities.

The light curve properties and host galaxy photometry of AT 2020mot indicate a central black hole mass of $M_{\text{BH}} \approx 3-6 \times 10^{6} M_{\odot}$, and stream shocks and accretion disk reprocessing models both find masses that are within one another's range of error. Relations between host galaxy properties and black hole mass are historically informed by larger-mass black holes, and thus predict larger black holes as well. However, nearly all our results are consistent with a mass that is $< 10^{7} M_{\odot}$.

We find that the unique $i$-band signatures may be explained by models of two concentric rings of face-on dust reprocessing the TDE emission into the infrared. These modeled rings are inferred at distances of 0.001pc and 0.09pc away from the central flare, which are among the smallest ever reported for the proximity of dust to a supermassive black hole. However, the the anomalous $i$-band excess never seen in another TDE may be due to another reddening effect such as Extended Red Emission. 

We have shown the potential of TDEs to probe a new regime of dust around galaxy centers. These results are independent of the disputed TDE emission source and hence pose a robust use of TDEs. NIR observations are often excluded from high-cadence rapid followup of TDEs in favor of early-time X-ray and UV observations to constrain the flaring mechanism. However, the inclusion of the $i$-band not only may probe different dust grain types and sizes than previously assumed for galactic nuclei, but in general, any IR observations achieved concurrent with the optical peak and decline may illuminate the closest surviving dust to SMBHs. We encourage systematic analysis of multi-band optical light curves of TDEs with the inclusion of the $i$-band or comparable near-infrared bands in high-cadence light curves, such as with ZTF samples, to explore the evidence of dust echoes in other sources, especially in cases with WISE light curves which indicate a dust echo as well. We similarly encourage follow-up of ongoing and new interesting nuclear transients with near-infrared photometry in high-cadence optical surveys to probe a new regime of dust echo signatures.

\vspace{1cm}
M.N. is supported by the NSF Graduate Research Fellowship Program. D.A.H., E.P.G., and C.P. are supported by NSF grant Nos. AST-1911225 and AST-1911151. I.A. is a CIFAR Azrieli Global Scholar in the Gravity and the Extreme Universe Program and acknowledges support from that program, from the European Research Council (ERC) under the European Unions Horizon 2020 research and innovation program (grant No. 852097), from the Israel Science Foundation (grant No. 2752/19), from the United States—Israel Binational Science Foundation (BSF), and from the Israeli Council for Higher Education Alon Fellowship. This research makes use of observations from the Las Cumbres Observatory global telescope network as well as the NASA/IPAC Extragalactic Database (NED), which is operated by the Jet Propulsion Laboratory, California Institute of Technology, under contract with NASA.

%

\vspace{5mm}
\facilities{LCO, ZTF, Swift(XRT and UVOT), WISE}


\software{astropy \citep{2013A&A...558A..33A,2018AJ....156..123A}, lcogtsnpipe \citep{Valenti16}, }



\pagebreak

\appendix

\section{Filter Differences between ZTF and Las Cumbres}

\begin{figure}
    \centering
    \includegraphics[scale=0.5]{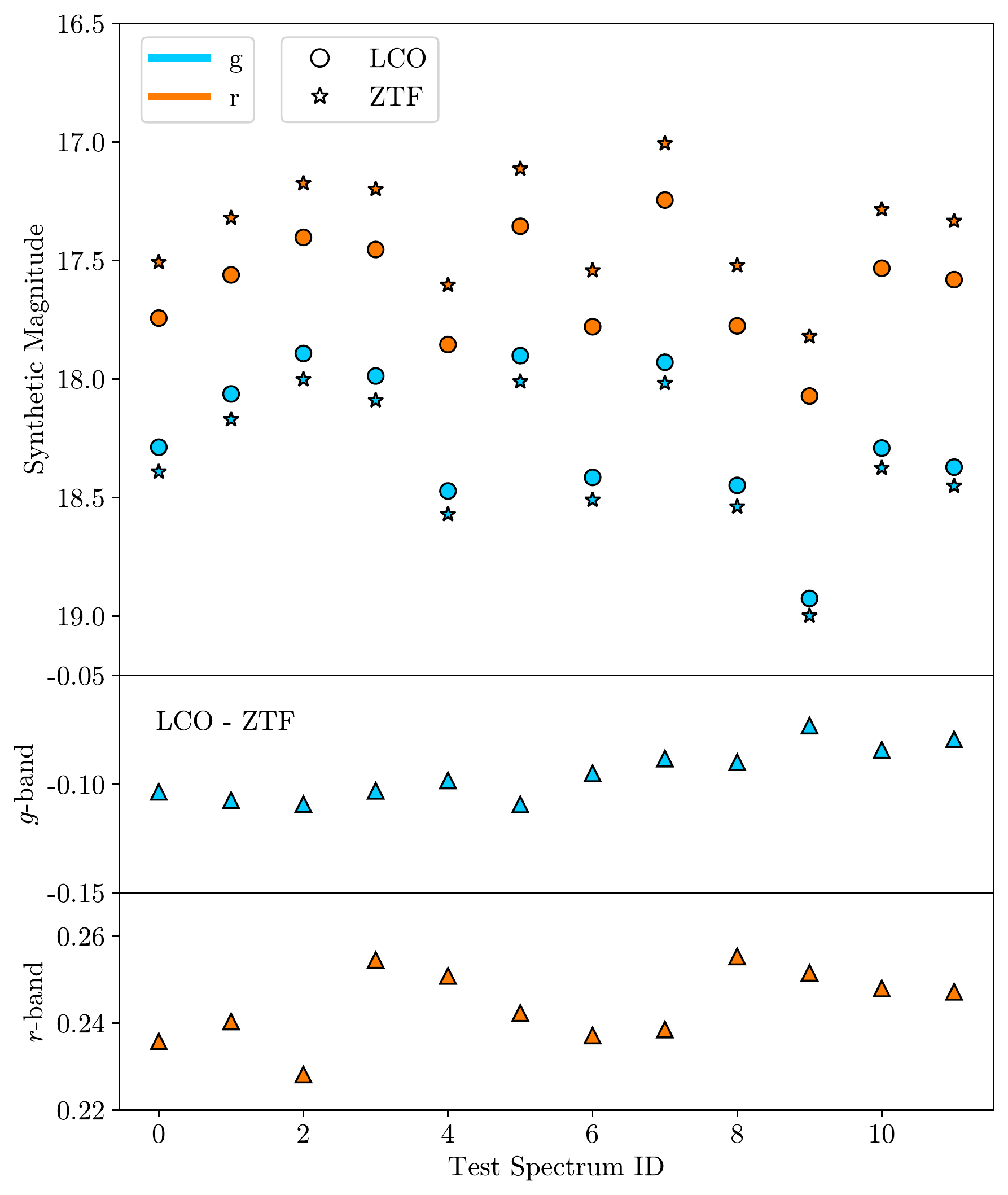}
    \caption{The synthetic photometry of 12 test spectra from FLOYDS-N when using Las Cumbres filters (filled circles) and ZTF filters (filled stars). The top panel shows both $g$ (blue) and $r$ photometry, the middle panel shows the difference between the magnitudes measured by LCO and ZTF in the $g$ band, and the bottom panel shows that of the $r$ band. The ZTF $r$-band synthetic photometry is an average of 0.25mag brighter than that of the Las Cumbres $r$-band filter.}
    \label{fig:ZTF_LCO_diffs}
\end{figure}

Although the Zwicky Transient Facility (ZTF) and Las Cumbres Observatory both use filters described as $g$ and $r$ bands which are often analyzed simultaneously in transient reports, we find a significant difference between the ZTF and Las Cumbres $r$-band filters that warrants their individual analyses. In Figure \ref{fig:ZTF_LCO_diffs} we show the synthetic photometry of 12 test spectra, each taken from Las Cumbres Observatory's 2m telescope. We find that the ZTF $g$-band magnitudes derived from spectra are systematically dimmer than those of Las Cumbres' $g$-band by an average 0.1 mag; by contrast, the ZTF $r$-band derives synthetic photometry that is an average of 0.25mag brighter than that of the Las Cumbres $r$-band filter. These differences produce light curves that may appear to have a wide scatter in $r$-band observations when the sensitivity curves of each filter are not properly taken into account, simultaneously affecting estimates in blackbody fitting and light curve fitting with tools such as \texttt{MOSFiT} \citep{Guillochon2017}.

\section{All Reduced Photometry}

We report all photometry that was reduced and subtracted for this work from Las Cumbres in Table \ref{table:LCOphot} and from Swift in Table \ref{table:UVphot}.

\begin{deluxetable}{cccccc}[t!]
\label{table:LCOphot}
    \tablehead{
    \colhead{MJD} & \colhead{$B$} & \colhead{$g$} & \colhead{$V$} & \colhead{$r$} & \colhead{$i$}
    }
    \startdata
59078.6 & 18.30 $\pm$ 0.01 & 18.21 $\pm$ 0.03 & 18.25 $\pm$ 0.01 & ... & ... \\
59084.5 & ... & 18.32 $\pm$ 0.02 & ... & 18.50 $\pm$ 0.03 & 18.35 $\pm$ 0.05 \\
59089.5 & 18.47 $\pm$ 0.02 & 18.30 $\pm$ 0.01 & 18.35 $\pm$ 0.02 & 18.49 $\pm$ 0.02 & 18.36 $\pm$ 0.03 \\
59099.4 & 18.62 $\pm$ 0.02 & ... & 18.57 $\pm$ 0.02 & 18.55 $\pm$ 0.03 & 18.33 $\pm$ 0.03 \\
59108.4 & 18.65 $\pm$ 0.02 & 18.57 $\pm$ 0.02 & 18.61 $\pm$ 0.03 & 18.75 $\pm$ 0.03 & 18.45 $\pm$ 0.04 \\
59113.2 & 18.73 $\pm$ 0.03 & 18.66 $\pm$ 0.02 & 18.56 $\pm$ 0.02 & 18.72 $\pm$ 0.03 & 18.56 $\pm$ 0.05 \\
59121.4 & 18.90 $\pm$ 0.03 & 18.79 $\pm$ 0.03 & 18.63 $\pm$ 0.03 & 18.92 $\pm$ 0.04 & 18.67 $\pm$ 0.05 \\
59128.4 & 18.99 $\pm$ 0.02 & 18.67 $\pm$ 0.03 & 18.91 $\pm$ 0.03 & 19.02 $\pm$ 0.05 & 18.68 $\pm$ 0.05 \\
59140.4 & 19.19 $\pm$ 0.03 & 19.08 $\pm$ 0.02 & 19.09 $\pm$ 0.04 & 19.16 $\pm$ 0.04 & 18.95 $\pm$ 0.05 \\
59145.2 & 19.32 $\pm$ 0.03 & 19.07 $\pm$ 0.02 & 19.25 $\pm$ 0.03 & 19.26 $\pm$ 0.04 & 18.94 $\pm$ 0.04 \\
59155.3 & 19.64 $\pm$ 0.09 & ... & 19.36 $\pm$ 0.06 & ... & ... \\
59160.2 & 19.63 $\pm$ 0.04 & 19.38 $\pm$ 0.02 & 19.44 $\pm$ 0.06 & 19.43 $\pm$ 0.05 & 19.06 $\pm$ 0.05 \\
59165.3 & 19.78 $\pm$ 0.04 & 19.48 $\pm$ 0.03 & 19.61 $\pm$ 0.06 & 19.51 $\pm$ 0.04 & 19.04 $\pm$ 0.04 \\
59173.2 & 19.91 $\pm$ 0.05 & 19.48 $\pm$ 0.03 & 19.79 $\pm$ 0.04 & 19.57 $\pm$ 0.05 & 19.02 $\pm$ 0.04 \\
59178.3 & 20.07 $\pm$ 0.07 & 19.62 $\pm$ 0.05 & 19.76 $\pm$ 0.06 & 19.64 $\pm$ 0.10 & ... \\
59202.2 & ... & 19.82 $\pm$ 0.05 & 19.99 $\pm$ 0.09 & 19.70 $\pm$ 0.09 & 19.35 $\pm$ 0.08 \\
59222.2 & ... & 20.13 $\pm$ 0.05 & ... & ... & ... \\
59227.2 & ... & 20.36 $\pm$ 0.08 & ... & ... & 19.54 $\pm$ 0.09 \\
    \enddata
    \caption{Optical photometry from the Las Cumbres Observatory, reduced and subtracted with \texttt{lcogtsnpipe}. $BV$ magnitudes are given in the Vega system while $gri$ magnitudes are given in the AB system.}
    \label{tab:Lco_phot}
\end{deluxetable}

\begin{deluxetable}{ccccc}[t!]
\label{table:UVphot}
    \tablehead{
    \colhead{MJD} & \colhead{$UVW2$} & \colhead{$UVW1$} & \colhead{$UVM2$} & \colhead{$U$}
    }
    \startdata
59075.1 & 16.74 $\pm$ 0.03 & 16.80 $\pm$ 0.03 & 16.91 $\pm$ 0.03 & 17.19 $\pm$ 0.04 \\
59078.7 & 16.79 $\pm$ 0.10 & 16.79 $\pm$ 0.11 & 16.91 $\pm$ 0.11 & 17.22 $\pm$ 0.13 \\
59082.6 & 16.83 $\pm$ 0.09 & 0.00 $\pm$ 0.00 & 16.95 $\pm$ 0.09 & 17.30 $\pm$ 0.11 \\
59089.8 & 16.84 $\pm$ 0.09 & 16.95 $\pm$ 0.10 & 17.04 $\pm$ 0.10 & 17.50 $\pm$ 0.12 \\
59096.1 & 16.94 $\pm$ 0.10 & 17.14 $\pm$ 0.09 & 17.10 $\pm$ 0.09 & 17.57 $\pm$ 0.12 \\
59103.6 & 17.14 $\pm$ 0.10 & 17.07 $\pm$ 0.09 & 17.35 $\pm$ 0.10 & 0.00 $\pm$ 0.00 \\
59123.6 & 17.52 $\pm$ 0.10 & 17.41 $\pm$ 0.11 & 17.48 $\pm$ 0.11 & 17.75 $\pm$ 0.13 \\
59130.3 & 17.53 $\pm$ 0.10 & 17.67 $\pm$ 0.11 & 17.67 $\pm$ 0.11 & 17.97 $\pm$ 0.14 \\
59143.6 & 17.80 $\pm$ 0.07 & 18.04 $\pm$ 0.10 & 18.03 $\pm$ 0.10 & 0.00 $\pm$ 0.00 \\
59151.5 & 18.08 $\pm$ 0.12 & 0.00 $\pm$ 0.00 & 0.00 $\pm$ 0.00 & 0.00 $\pm$ 0.00 \\
    \enddata
    \caption{Ultraviolet photometry from Swift-UVOT, reduced and subtracted with the \texttt{heasarc} pipeline. Magnitudes are given in the Vega system.}
    \label{tab:swiftphot}
\end{deluxetable}

\subsection{Host Photometry Analysis}

\texttt{BAGPIPES} \citep{Carnall2018} generates model spectra and fits to observed photometry of a galaxy to calculate the probability distribution functions (PDF) of values relevant to the star formation history (SFH), dust, and metallicity. We start with the same choice of SFH and priors as in \cite{Carnall2018} for our initial fit. We use the stellar synthesis models as outlined in \cite{BC03} and updated in 2016\footnote{\url{http://www.bruzual.org/~gbruzual/bc03/Updated_version_2016/}} assuming a \cite{Chabrier2003} initial mass function. The star formation history is modeled as a double-power law,
\begin{equation}
    \text{SFR}(t) \propto \left [ \left (\frac{t}{\tau} \right )^{\alpha}  \left (\frac{t}{\tau} \right )^{-\beta} \right ]^{-1}
\end{equation}
where $\alpha$ and $\beta$ are the rising and falling slopes respectively, while $\tau$ is related to the time of peak star formation. Our priors on $\alpha$ and $\beta$ are logarithmically uniform and range from 0.01 to 1000.0. We fit for metallicity with a uniform prior between 0.0 and 2.5 times solar metallicity, and we use a \cite{Calzetti2000} dust law with a uniform prior for extinction running from $A_{V} = (0.0, 4.0)$. We exclude a nebular component as the host spectrum does not show emission features. We also fix the redshift to the spectroscopically determined value $z=0.07$.

\begin{figure}[t]
    \centering
    \includegraphics[scale=0.33]{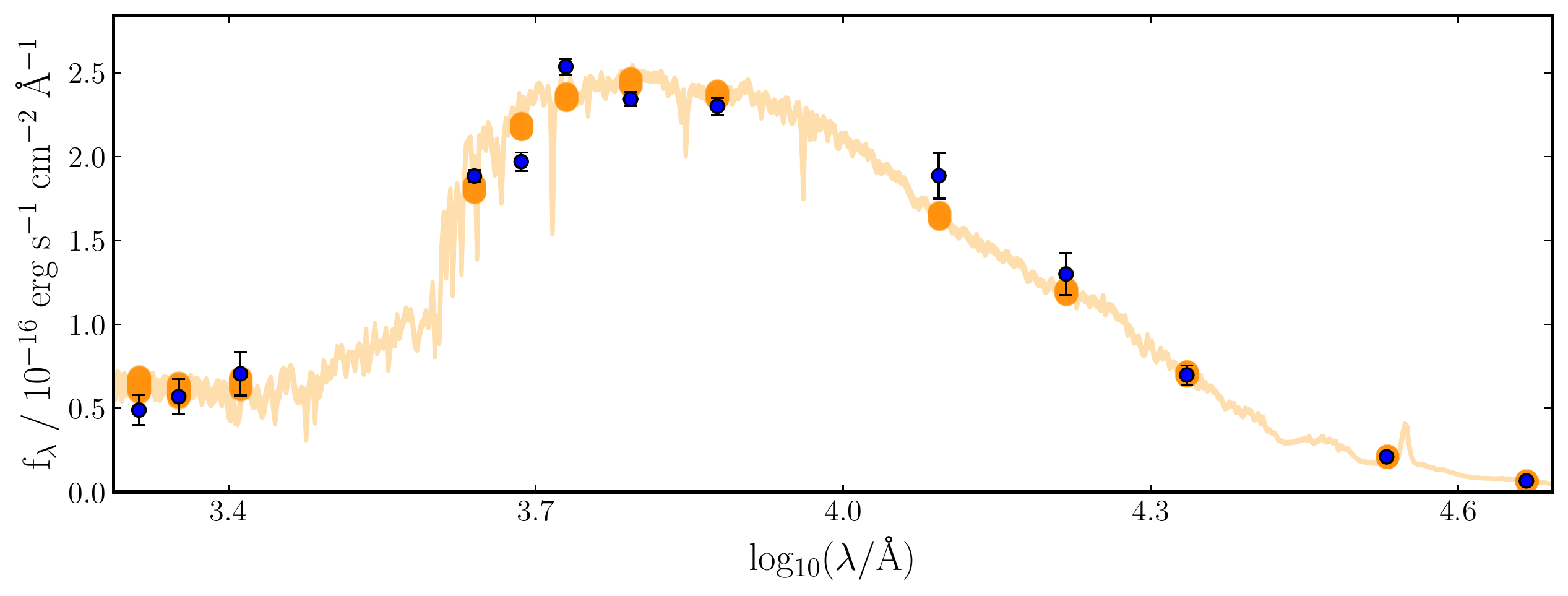}
    \caption{\texttt{BAGPIPES} fit to the J003113.52+850031.8 SED.}
    \label{fig:galaxyfit}
\end{figure}
\begin{figure}[t]
    \centering
    \includegraphics[scale=0.3]{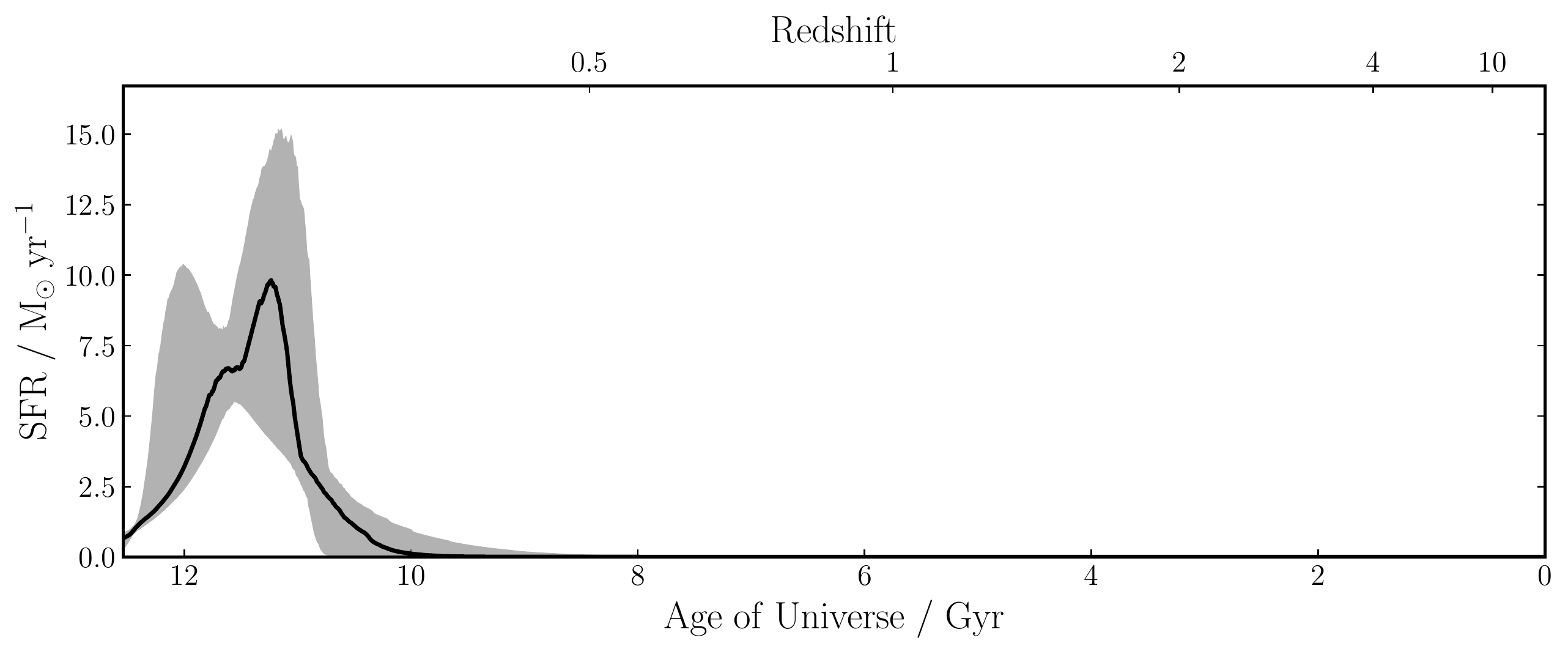}
    \caption{\texttt{BAGPIPES} best-fit star formation history to J003113.52+850031.8 SED.}
    \label{fig:galaxysfh}
\end{figure}


\bibliographystyle{aasjournal}
\bibliography{main}{}



\end{document}